\documentclass[12pt]{article}
\usepackage{graphicx, amsmath, amssymb, cite, setspace, color, relsize, bm, bbold, array, hyperref, dsfont, bbold, xcolor, cancel,soul,placeins,hyperref,amsfonts,pifont}

\usepackage[top=1 in, bottom=1 in, left=.9 in, right=.9 in]{geometry}

\newcommand\To{\rule{0pt}{4.5ex}}       
\newcommand\Bo{\rule[-3.0ex]{0pt}{0pt}} 

\newcommand{\captionfonts}{\footnotesize} 
\makeatletter
\long\def\@makecaption#1#2{%
  \vskip\abovecaptionskip
  \sbox\@tempboxa{{\captionfonts #1: #2}}%
  \ifdim \wd\@tempboxa >\hsize
    {\captionfonts #1: #2\par}
  \else
    \hbox to\hsize{\hfil\box\@tempboxa\hfil}%
  \fi
  \vskip\belowcaptionskip}
\makeatother

\def\lsim{ \lower .75ex \hbox{$\sim$} \llap{\raise .27ex
\hbox{$<$}} }
\def\gsim{ \lower .75ex \hbox{$\sim$} \llap{\raise .27ex
\hbox{$>$}} } 
\def\bn{\bigskip \noindent}

\renewcommand{\title}[1]{\vbox{\center\LARGE{#1}}\vspace{5mm}}
\renewcommand{\author}[1]{\vbox{\center#1}\vspace{5mm}}
\newcommand{\address}[1]{\vbox{\center\em#1}}

\newcommand{\starttext}{
\setcounter{footnote}{0}
\renewcommand{\thefootnote}{\arabic{footnote}}}

\newcommand{\be}{\begin{equation}}
\newcommand{\bea}{\begin{eqnarray}}
\newcommand{\eea}{\end{eqnarray}}
\newcommand{\beq}{\begin{equation}}
\newcommand{\ee}{\end{equation}}

\let\oldsqrt\sqrt
\def\sqrt{\mathpalette\DHLhksqrt}
\def\DHLhksqrt#1#2{%
\setbox0=\hbox{$#1\oldsqrt{#2\,}$}\dimen0=\ht0
\advance\dimen0-0.2\ht0
\setbox2=\hbox{\vrule height\ht0 depth -\dimen0}%
{\box0\lower0.4pt\box2}}

\def\sc{\setcounter{equation}{0}}

\begin{document}
\begin{titlepage}

\rightline{}
\bigskip
\bigskip\bigskip\bigskip\bigskip
\bigskip

\centerline{\Large \bf {The Complexity Geometry of a Single Qubit}}

\bn

\bigskip

\bigskip
\begin{center}

\author{Adam R. Brown and Leonard Susskind}

\address{Google, Mountain View, CA 94043, USA}

\address{Stanford Institute for Theoretical Physics and Department of Physics, \\
Stanford University, Stanford, CA 94305, USA}

\end{center}

\begin{center}
\bf     \rm

\bigskip

\end{center}

\begin{abstract}

 The computational complexity of a quantum state quantifies how hard it is to make. `Complexity geometry', first proposed by Nielsen, is an approach to defining computational  complexity using the tools of differential geometry. Here we demonstrate many of the attractive features of complexity geometry using the example of a single qubit, which turns out to be rich enough to be illustrative but simple enough to be illuminating.

\medskip
\noindent
\end{abstract}

\let\thefootnote\relax\footnotetext{email: \tt{mr.adam.brown@gmail.com}}

\end{titlepage}

\starttext \baselineskip=17.63pt \setcounter{footnote}{0}

\vfill\eject

\tableofcontents

\vfill\eject

\section{Introduction}

The inner-product distance gives a metric on quantum states. Two states that have a large inner product are close; two states that have a small inner product are far; two states that are orthogonal are as distant as any two states can be.

But there are some intuitive notions of proximity that are not captured by the inner product.  In the example below, we consider the following two states:  the Earth plus a single spin-up electron; and the Earth plus a single spin-down electron. Since these two states are orthogonal, the inner-product distance says they are maximally separated. Nevertheless, since they differ by only a single spin, there is a sense in which they are close. The intuition is that they are close since it is `easy' to change one into the other---you just need to flip a single spin. 

\begin{figure}[htbp] 
   \centering
   \includegraphics[width=5in]{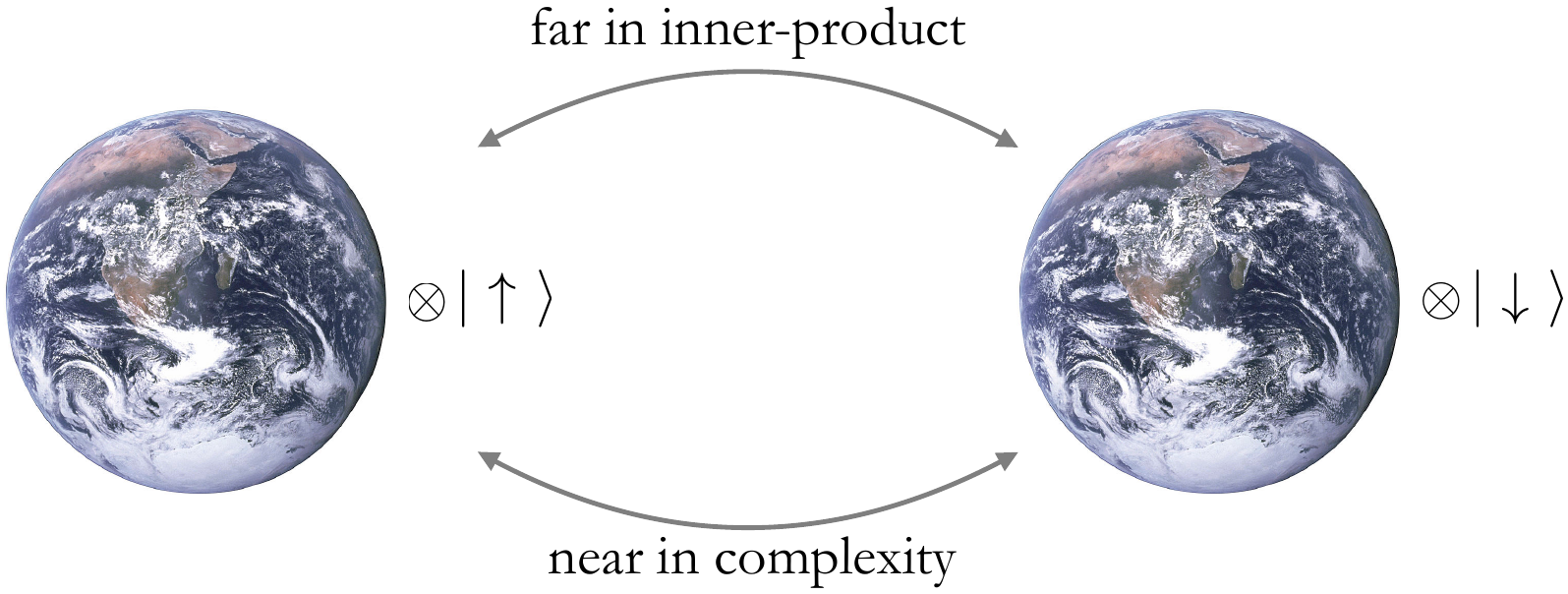} 
   \caption{According to the inner-product metric, these two states are as far apart as any two states can be: they are orthogonal. The `complexity distance' captures the sense in which they are nevertheless close.}
   \label{fig-NearEarth}
\end{figure}

We'd like a notion of proximity that captures the sense in which the two states of Fig.~\ref{fig-NearEarth} are close.  Computational complexity provides such a notion. The relative computational complexity of two quantum states quantifies how hard it is, given one, to make the other. Since the two Earths are identical, the complexity distance is small; if the two Earths had very different weather patterns, the complexity distance would be large.

This paper will study one approach to characterizing the complexity distance, which will involve making mathematically precise our notion of when a transformation is easy. 

Of course, whether a given task is `easy' or `hard' depends on one's capabilities. Questions of quantum computational complexity were first addressed by those who wanted to know what we would be able to do with a quantum computer. In this context, the primitive capability would be the ability to lay out gates in a quantum circuit. Thus the traditional definition of complexity, which we will review in Sec.~\ref{sec:gatecomplexity}, is the gate definition that asks how many gates are needed in the smallest quantum circuit that implements the desired transformation. 

The gate definition of complexity is well-defined, carries the weight of tradition and the benefits of incumbency, and is a natural measure from the point of view of electrical engineers planning to build actual quantum circuits. But for some purposes it has a number of drawbacks, which we will describe in Sec.~\ref{sec:gatecomplexity}, not least that it is discontinuous.

In this paper we will instead explore another idea, due to Nielsen and collaborators \cite{Nielsen1,Nielsen2,Nielsen3,Nielsen4,NielsenSingleQubit}, that seeks to replace the discrete counting of quantum gates with a continuous definition of complexity. As we will explain in Sec.~\ref{sec:complexitygeometry}, the complexity geometry approach puts a new metric on Hilbert space, different from the inner-product metric. Unlike the inner-product metric, this new metric doesn't treat all directions in the tangent space equally. Instead it stretches directions that are hard to move in, assigning them a large distance. This definition then permits us to bring all the tools of differential geometry to bear on the question of complexity. \\

Our background motivation is twofold. First, we wish to better understand the holographic complexity conjecture \cite{Susskind:2014rva,Stanford:2014jda,Brown:2015bva,Brown:2015lvg}, and therefore better understand the emergence of spacetime in quantum gravity. If this conjecture is to be made more precise, we will need an exact definition of the complexity of the holographic field theory. Such a definition could potentially be provided by complexity geometry, and our first background motivation will be to evaluate it for that role; as we will see, it is in many ways a better candidate than the definition provided by gate-counting. Our second background motivation is to explore whether the tantalizing analogies between quantum complexity and statistical thermodynamics \cite{Brown:2017jil} could be better teased out with a more amenable definition of complexity.

While those are our motivations, the direct goal of this paper will be more modest. 
The complexity geometry, while simple to define, can for large systems become extremely unwieldy.  To try to build our intuitions, and to demonstrate the power of this way of thinking in the simplest non-trivial context, we will in this paper calculate the complexity geometry of a single qubit. We will see that the complexity geometry of a single qubit is simple enough to be intuitive, and to make illuminating contact with the rotation of rigid bodies, but complicated enough to exhibit many of the signature phenomena of multi-qubit complexity geometry. \\

`Complexity Geometry' should not be confused with `Geometric Complexity'  \cite{GeometricComplexity}, which as far as we know is completely unrelated. The complexity geometry of a holographic system is also not to be confused with the geometry of the holographic dual---they are not the same, and their dimensionalities exponentially differ.\\

In Sec.~\ref{sec:FunctionVSState} we will distinguish the two kinds of objects whose complexities we may wish to calculate: unitaries and states. In Sec.~\ref{sec:GateVSGeometry} we will discuss the two ways we might seek to characterize the complexity of these objects: gate counting and complexity geometry. \\

\subsection{Unitary \emph{vs}. state complexity} \label{sec:FunctionVSState}

In this paper we examine the complexity both of \emph{unitaries} and of \emph{pairs of states}. \\

\noindent {\bf Unitary complexity} is a property of a unitary transformation that quantifies how hard it is to implement. This is the quantum analogue of the classical function complexity. The complexity geometry of unitaries will provide a new metric on the unitary group. We will see that the metric is  homogeneous but not isotropic, and is right-invariant but not left-invariant. \\

\noindent We may also speak of the relative complexity of a \emph{pair} of unitaries, defined as $\mathcal{C}
[ U_1; U_2  ] \equiv \mathcal{C}
[ U_1 U_2^{-1}  ].$ For $U_2 = \mathds{1}$ this reduces to the standard definition of the unitary complexity of $U_1$. \\

\noindent  {\bf State complexity} is a property of a pair of states, and quantifies how hard it is, given one, to make the other. The complexity geometry of quantum states will provide a new metric on Hilbert space. We will see that this metric is not left-invariant, which means it is neither homogeneous nor isotropic.\\

These two kinds of complexity can be related. The state complexity is the complexity of the \emph{least} complex unitary that connects the reference and target states\footnote{Note that the converse would not be correct---unitary complexity is \emph{not} the complexity of the  \emph{most} complex pair of states that the unitary connects. Indeed this wouldn't be true even if we replaced `complexity distance' with `inner-product distance', since we can have $\langle \psi | U | \psi \rangle = 0$ while $\textrm{Tr}[U]$ is still close to maximal.},
\begin{equation}
\mathcal{C}_\textrm{state}[ |\psi_1 \rangle ; |\psi_0 \rangle  ] = \textrm{min}_U \mathcal{C}_\textrm{unitary}[U] \ \ \textrm{where} \ \  |\psi_1 \rangle = U |\psi_0 \rangle. \label{eq:stateintermsoffunction}
\end{equation}

\vspace{5mm}

While real life is replete with examples of things that are easier to do than to undo, in the field of computational complexity it is conventional to choose a definition such that $\mathcal{C}[U] = \mathcal{C}[U^{-1}]$. This has the consequence that relative complexity is symmetric: the distance from $A$ to $B$ is equal to the distance from $B$ to $A$. In the context of gate complexity this will mean that if $g$ is a fundamental gate, then so too is $g^{-1}$; in the context of complexity geometry this will mean that we will be able to use a metric that is symmetric:  $\mathcal{C}[ U_1; U_2  ] \equiv \mathcal{C}[ U_1 U_2^{-1}  ] = \mathcal{C}[ U_2; U_1  ] \equiv \mathcal{C}[ U_2 U_1^{-1}  ] $.

 \subsection{Gate complexity \emph{vs}. complexity geometry} \label{sec:GateVSGeometry}

The `complexity' of a transformation characterizes how hard it is to implement. This paper will explore the `complexity geometry' definition of hardness, but as a foil let's first briefly describe another popular way to characterize hardness: gate complexity. 

\subsubsection{Gate Complexity} \label{sec:gatecomplexity}

The gate definition of complexity is a natural one for those trying to actually build quantum circuits out of component gates. It tells you how many of those components you will need: 
\begin{eqnarray}
\textrm{\tt `Gate Complexity'} & \equiv & \textrm{the number of primitive gates in the smallest quantum} \ \ \   \\  && \textrm{circuit that implements the transformation (to within $\epsilon$) \nonumber}
\end{eqnarray}
To finish unambiguously specifying the definition, we will further need to:
\begin{itemize}
\item Choose the set of primitive gates.

For example, a classic choice for the set of primitive gates out of which we will build our circuits is to use the two-qubit `CNOT' gate, together with the one-qubit gates `Hadamard' and `$\pi/8$' (otherwise known as the `T' gate; see e.g. \cite{Nielsen:2011:QCQ:1972505} for the definition of these gates, and more discussion of gate complexity). Together these three primitive gates are `universal', in the sense that by compiling these gates we can approximate any $N$-qubit unitary with arbitrary accuracy.

(Were we just concerned with making all the $1$-qubit unitaries, we could drop the CNOT.)
\item Choose the `tolerance' $\epsilon$ with which we wish to approximate our desired unitary. 

Since there are uncountably many $N$-qubit unitaries but by assumption only countably many primitive gates, we cannot hope to be exact.  Instead we settle for making a close enough approximation. For example, we might declare that a circuit $U_\textrm{circuit}$ is good enough if 
\begin{equation}
1 - 
|\textrm{Tr}[U_\textrm{circuit}^\dagger U_\textrm{target}]| < \epsilon = 10^{-6}.
\end{equation}
(Here and throughout this paper we normalize the trace such that Tr$\mathds{1} = 1$.)
\end{itemize}

\begin{figure}[htbp] 
   \centering
   \includegraphics[width=4in]{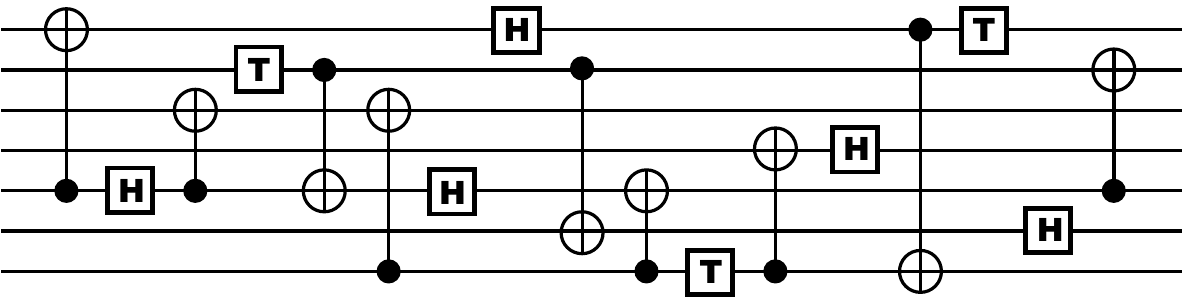} 
   \caption{The `gate' definition of complexity imagines implementing $U$ by building a quantum circuit out of primitive gates. In this example, the primitive gates each act only on one or two qubits at a time, but in combination they may approximate any $N$-qubit unitary with arbitrary accuracy. The computational complexity is then defined as the total number of gates in the smallest circuit that approximates the desired unitary.}
   \label{fig-quantumcircuit}
\end{figure}

\paragraph{Drawbacks of Gate Complexity.} The gate-counting definition of complexity has a number of  features that for certain applications are undesirable:

\begin{enumerate}
\item The choice of primitive gates seems arbitrary.

Why those particular one-qubit and two-qubit primitive gates, out of all possible gates?

\item The definition requires, and is sensitive to, the introduction of an arbitrary `tolerance'. 

Why $\epsilon = 10^{-6}$, rather than some other value?

\item The complexity is discontinuous. 

Two points in the unitary group (or in Hilbert space) can be arbitrarily close in inner product but exponentially far in complexity. 
\end{enumerate}

Some of these drawbacks could be addressed by tinkering with our definition of gate complexity. For example, we could give ourselves as primitive gates all possible two-qubit unitaries.
This would largely address complaint \#1 (leaving over only the question of why we permitted two-qubit gates but not three-or-more-qubit gates); and it would mean that we could entirely dispense with a tolerance---with this continuous set of primitives, using O($4^N$) gates we can hit everywhere in SU($2^N$) \emph{exactly} \cite{Nielsen:2011:QCQ:1972505}. Nevertheless, this definition would still give complexities that are discontinuous---indeed if each gate still had unit cost there would be pairs of states that are arbitrarily close in inner product but that have exponentially large relative gate complexity, and the set of states with sub-maximal complexity would be measure zero.

Because of these limitations, many computer scientists try to look only at coarse-grained complexity classes that are insensitive to these definitional ambiguities. For example, the complexity class BQP (the class of (decision) problems that can be probabilistically solved in polynomial time using a quantum computer) is so broad as to be largely choice-independent.

However, it is somewhat disappointing if that is the best we can do. We would like to have a definition that assigns a  quantitative number to the complexity of a particular unitary and have that number be somewhat robust against arbitrary choices, perhaps up to a multiplicative renormalization of the complexity.  With this motivation, we turn to complexity geometry. 

\subsubsection{Complexity geometry} \label{sec:complexitygeometry} 

The `complexity geometry' approach to quantifying quantum complexity was invented by Michael Nielsen and collaborators \cite{Nielsen1,Nielsen2,Nielsen3,Nielsen4,NielsenSingleQubit}. Rather than leaping through Hilbert space with discrete gates, instead we imagine smoothly flowing from state to state along continuous paths. The length of the shortest path gives the computational complexity. 

Were it equally easy to move in every direction in Hilbert space, this definition would recover the standard inner-product metric. But not all directions \emph{are} equally hard: the whole point of inventing the notion of `complexity' is that the inner-product metric does not accurately reflect how difficult a transformation is to implement. Instead, the complexity metric reflects that some transformations are hard to implement (i.e. some directions in Hilbert space are hard to move in) by `stretching' the metric in those directions. By assigning hard directions a large metric penalty, even steps in hard directions that are quite short in the inner-product metric become far in the complexity metric. 

The motivation of the originators of the complexity geometry idea seems to have been primarily to use it as a tool to bound the value of the gate complexity. Our motivation is somewhat different. Instead of using it to better understand the gate definition of complexity, we consider the geometric definition of complexity to be an interesting quantity in its own right, with a potentially interesting holographic dual. We will study the complexity geometry as a candidate for being the fundamental definition of complexity.  Thus we will not think of complexity geometry as being a continuous approximation to gate complexity, but rather think of gate complexity as being a discrete approximation to complexity geometry. \\

For the rest of this paper we will explore the simple case of the complexity geometry of a single qubit, but first let's give the general mathematical definition of the complexity geometry for a general number of qubits $N$. We'll start by defining the \emph{unitary} complexity geometry; the \emph{state} complexity geometry then follows by using Eq.~\ref{eq:stateintermsoffunction}.

Following Nielsen, the unitary group we will consider is SU($2^N$). Up to a discrete  identification, this is the group of purity-preserving transformations of $N$ qubits that ignore the global phase. The tangent space of the unitary group is spanned by Hermitians. For a single qubit, a complete basis for the tangent space of 
 SU(2)  is provided by the $2^2 -1 = 3$ Pauli matrices, $\sigma_i$. For $N$ qubits, a complete basis for the tangent space of SU($2^N$)  is provided by the $2^{2N} - 1$ `generalized' Pauli matrices, $\sigma_I$. The generalized Pauli matrices are composed of direct products of single-qubit Pauli matrices and identity operators. Thus a typical example might be
\begin{equation}
\textrm{example of generalized Pauli }\sigma_I  = \sigma_{1x} \otimes \sigma_{2z} \otimes \mathds{1}_3 \otimes \ldots \otimes \sigma_{Nz} \ . \label{eq:examplePauli}
\end{equation}

The inner-product metric on SU($2^N$) gives the inner-product distance, $ds$, between two nearby unitaries, $U$ and $U + d U$, as\footnote{Recall that if given a vector $\vec{V}$ and a complete orthonormal basis $\{ \vec{e}_i \}$, the inner-product can be written either as $\vec{V} \cdot \vec{V}$ or as $\sum_i (\vec{V} \cdot \vec{e}_i)^2$; Eqs.~\ref{eq:generalNinnerproductmetric}-\ref{eq:manifestleft} reflect that we have the same freedom for matrices. Recall also that we have normalized the trace such that Tr[$\mathds{1}] = 1$.}
\begin{eqnarray}
ds^2 \Bigl|_\textrm{inner-product} &=& \textrm{Tr}[dU^\dagger dU ]  \label{eq:generalNinnerproductmetric} \\
&=&     \sum_{IJ} \textrm{Tr}[i dU U^\dagger  \sigma_I] \delta_{IJ} \textrm{Tr}[i dU U^\dagger  \sigma_J]  \label{eq:manifestright}\\
&=&   \sum_{IJ} \textrm{Tr}[i U^\dagger dU  \sigma_I] \delta_{IJ} \textrm{Tr}[i U^\dagger dU   \sigma_J] 
\label{eq:manifestleft} . 
\end{eqnarray}
(The form in Eq.~\ref{eq:manifestright} is manifestly right-invariant $U \rightarrow U U_R$; the form in Eq.~\ref{eq:manifestleft} is manifestly left-invariant $U \rightarrow U_L U$; the form in Eq.~\ref{eq:generalNinnerproductmetric} makes both invariances manifest. Since the inner-product metric is invariant under $U \rightarrow U_L U U_R$ it is sometimes called the `bi-invariant' metric.) The inner-product metric treats all the tangent directions $\sigma_I$ the same, as is reflected by the coefficient of every diagonal term in Eq.~\ref{eq:manifestright} being identical. 

The complexity metric generalizes $\delta_{IJ}$ in Eq.~\ref{eq:manifestright} to a symmetric positive-definite penalty factor $\mathcal{I}_{IJ}$. The complexity distance, $ds$, between two nearby unitaries, $U$ and $U + d U$, is then given by 
\begin{equation}
ds^2\Bigl|_\textrm{complexity} = \sum_{IJ} \textrm{Tr}[i dU \, U^\dagger  \sigma_I] \mathcal{I}_{IJ} \textrm{Tr}[i dU \, U^\dagger  \sigma_J].
\end{equation}
While still right-invariant, this metric is no longer left-invariant, as will be discussed in Sec.~\ref{sec:leftandright}. 

Some directions may be harder to move in than others. This will be captured by the tensor $\mathcal{I}_{IJ}$, which penalizes `hard' directions with a larger coefficient. For example, a classic choice would be that the more qubits a certain $\sigma_I$ touches, the higher the penalty $\mathcal{I}_{II}$ it should be assigned. This reflects the fact that in the laboratory it is typically difficult to perform controlled operations that touch many qubits simultaneously\footnote{For example, the next generation of superconducting-qubit circuits are projected to have an error rate for two-qubit gates that is five times larger than the error rate for one-qubit gates \cite{GoogleQCGroup}.}. In \cite{Brown:2017jil}, we advocated for the proposal that the penalty factor should grow exponentially with the number of qubits the operator touches simultaneously (known as the operator's `weight'). 
We were motivated by two considerations. First, the complexity of an operator of fixed weight should be independent of the total number of qubits. Second, to reflect that the maximal complexity of a unitary can be exponentially large, the largest penalty factor also needs to be (at least) exponentially large.

For large $N$, this geometry is negatively curved and very complicated. Both the negative curvature and the unsimplicity of this metric are responsible for the chaotic behaviour of complexity, but they also make the analysis of the geometry difficult. It is therefore worth analyzing the simplest possible case. 

In this paper we will examine the complexity geometry of a single qubit, $N=1$, which turns out to be surprisingly rich.  In Sec.~\ref{sec:functionC} we'll look at the unitary complexity; in Sec.~\ref{sec:stateC} we'll look at the state complexity. In Sec.~\ref{sec:discussion}  we'll extract the general lessons that the one-qubit example can teach us about $N$-qubit complexity geometry.

\section{Unitary-Complexity Geometry of a Single Qubit} \label{sec:functionC}
\sc

In this section we will examine the transformations of a single pure-state qubit. Ignoring global phase, the group of linear transformations of a single qubit is, up to a discrete identification, SU(2). 
The standard inner-product metric on SU(2) is that of the round three-sphere\footnote{Again recall that we have normalized the trace so that Tr$[\sigma_i \sigma_j] = \delta_{ij}$.}
\begin{equation}
ds^2 = \textrm{Tr}[dU^\dagger dU ]  =       \textrm{Tr}[i d U U^\dagger \sigma_x]^2 +   \textrm{Tr}[i d U U^\dagger \sigma_y]^2 +  \textrm{Tr}[i d U U^\dagger \sigma_z]^2  . \label{eq:threespheremetric}
\end{equation}
But the complexity geometry will define a new metric on SU(2). 
Introducing a symmetric positive-definite two-index penalty factor $\mathcal{I}_{ij}$ gives
\begin{eqnarray}
 ds^2 &=& \textrm{Tr}[i d U U^\dagger \sigma_i] \mathcal{I}_{ij} \textrm{Tr}[i d U U^\dagger \sigma_j]   \label{squashedunitarysphere}
\\
& = &    \mathcal{I}_{xx}  \textrm{Tr}[i d U U^\dagger \sigma_x]^2 + \mathcal{I}_{yy}  \textrm{Tr}[i d U U^\dagger \sigma_y]^2 + \mathcal{I}_{zz}  \textrm{Tr}[i d U U^\dagger \sigma_z]^2  , \label{squashedsphere}
\end{eqnarray}
where in the last step we have adopted the basis in which $\mathcal{I}_{ij}$ is diagonal. In this section we will explore this metric. \\

In the typical application of complexity geometry as defined in Sec.~\ref{sec:complexitygeometry}, we imagined a metric that rewards $k$-locality (gates that touch only $k$ qubits at once are easy to implement) and punishes non-locality (gates that touch many qubits at once are given a large penalty factor). When there is only one qubit, such a distinction is meaningless: if there is only one qubit, no gate may touch more than a single qubit! If we treat all $1$-local transformations the same, the only possibly complexity metric on SU(2) is the standard round metric of Eq.~\ref{eq:threespheremetric}.

To reach Eq.~\ref{squashedsphere}, we had to do something different---we had to distinguish between  one-qubit operators. A complete basis for the tangent space of one-qubit operators is provided by the three Pauli matrices, which may be thought of as effecting rotations around the three Cartesian axes. The standard inner-product metric makes all three of these rotations equally easy. Instead, we imagined that some are easier than others. The three parameters $\mathcal{I}_{xx}, \ \mathcal{I}_{yy}, \ \& \ \mathcal{I}_{zz}$ capture how hard it is to rotate around the three different axes. 

There are any number of practical considerations that might give rise to unequal $\mathcal{I}_{xx}, \ \mathcal{I}_{yy}$,  \& $\mathcal{I}_{zz}$. For example, if our qubit is a spin, we could compile our unitary by applying a magnetic field according to some schedule $\vec{B}(t)$. Were it easy to apply a magnetic field in the $x$-$y$ plane, but for mechanical reasons hard to apply magnetic fields out of that plane, then this would be described by a small $\mathcal{I}_{xx}$ \& $\mathcal{I}_{yy}$ but large $\mathcal{I}_{zz}$. The complexity metric of Eq.~\ref{squashedsphere} assigns a degree of difficulty to a given schedule for compiling the unitary
\begin{equation}
\textrm{difficulty} = \int dt \sqrt{ \mathcal{I}_{xx} \, B_x(t)^2 + \mathcal{I}_{yy} \, B_y(t)^2 + \mathcal{I}_{zz} \, B_z(t)^2} .
\end{equation}
The complexity of the unitary is given by the least difficult schedule that produces it. 

 For many laboratory realizations of controlled qubits, there is indeed an asymmetry between the difficulty of implementing $\sigma_x$ or $\sigma_y$ and the difficulty of implementing $\sigma_z$. The asymmetry arises because the two quantum states in which the qubit is embodied are typically two non-degenerate energy levels, so the resting (`drift') evolution of the qubit naturally implements a $\sigma_z$ Hamiltonian, and not a $\sigma_x$ or $\sigma_y$ Hamiltonian.  Ironically this means that it is often harder to add additional $\sigma_z$ to the Hamiltonian (above and beyond the natural evolution) than it is to add $\sigma_x$ or $\sigma_y$. Ted White has explained to us that for the superconducting qubits being built by the Santa Barbara group,  $\sigma_x$ or $\sigma_y$ can be implemented with high fidelity using a properly shaped microwave pulse, whereas additional $\sigma_z$ requires the use of a harder-to-control and  lower-fidelity pseudo-DC pulse. Thus for these superconducting qubits it would be natural to take  $\mathcal{I}_{xx}$ \& $\mathcal{I}_{yy}$ small and equal but $\mathcal{I}_{zz}$ large. 

In this paper we will calculate the complexity geometry for a general single-qubit penalty factor, but will focus on the same case as is relevant for superconducting qubits---the case where one penalty factor is much larger than the other two, 
\begin{equation}
\textrm{penalty assignment focussed on in this paper:} \ \ \ \ \ \ \mathcal{I}_{zz} \gg \mathcal{I}_{xx} = \mathcal{I}_{yy}. \ \ \ \ \ \ \ \label{eq:preferredhierarchy}
\end{equation}
While this is the same hierarchy as pertains for superconducting qubits, that is not our reason for focusing on it. Instead, our motivations will become clear in Sec.~\ref{sec:discussion}. We will see that penalizing $\sigma_z$ more than $\sigma_x$ or $\sigma_y$ is the assignment that best captures the characteristic features of multi-qubit complexity geometry. 

Even though there is now only a single qubit, there is nevertheless a sense in which the hierarchy of Eq.~\ref{eq:preferredhierarchy} still fits into the scheme of taking higher-weight operators to be more penalized.  The qubit can be thought of as being comprised of two Majorana fermionic operators, which satisfy $\{ \psi_i , \psi_j \} = 2 \delta_{ij}$. The Hermitian operators $\psi_1$ and $\psi_2$ each have weight 1, whereas the Hermitian operator $i \psi_1 \psi_2$ has weight 2. On the other hand the commutation and anti-commutation relations of these three operators would be the same if we make the identification $\{ \psi_1, \psi_2, i \psi_1 \psi_2 \} \leftrightarrow \{ \sigma_x, \sigma_y, \sigma_z \}$. The penalty factor penalizes the weight-2 operator relative to the weight-1 operators. \\

To explore the complexity metric Eq.~\ref{squashedsphere} further, it will be helpful to introduce explicit coordinates. But first we will point out that we have seen this metric before. 

\subsection{Complexity geometry and the motion of rigid bodies} \label{sec:rigidbodies}
As well as representing the complexity geometry of a single qubit, the metric Eq.~\ref{squashedsphere} arises in a more mundane context: the rotation of a rigid body. 

\subsubsection{The motion of rigid bodies}
Recall that the configuration of a rigid body in three dimensions may be described by $R_{ij}$, the element of SO(3) that transforms from the lab frame to the body frame. For rigid bodies with symmetric moments of inertia, $\mathcal{I}_{ij} \sim \delta_{ij}$, it is equally easy to rotate around any axis, so the kinetic energy depends only on the angular velocity
\begin{equation}
\mathcal{I}_{ij} = \delta_{ij} \rightarrow \textrm{Kinetic Energy} = \frac{3}{2} \textrm{Tr}[\dot{R}^T \dot{R}]  = \frac{1}{2} \omega^T \omega,
\end{equation}
where $\omega_i \equiv \epsilon_{ijk}(R^T)_{jl} \dot{R}_{lk}$ and the factor of $3$ arises because of our trace normalization Tr$[R]\equiv \Sigma_i R_{ii}/3$. We may define the distance between any two configurations as the total angle through which you must rotate one to reach the other. The infinitesimal metric defined in this way is that of the round three-sphere,
\begin{equation}
\mathcal{I}_{ij} = \delta_{ij}  \rightarrow ds^2 = 3  \textrm{Tr}[ dR^{T}  dR ] .
\end{equation}
The geodesics of this metric both describe the locally extremal way of effecting a given transformation (the way to twist and turn the body so that it connects two configurations with the smallest total angular rotation), and also describes the path traced out through configuration space by the free motion of a symmetric rigid body.

Rigid bodies that are not spherically symmetric may have unequal moments of inertia. Consequently the kinetic energy required to rotate with a given angular speed is different for different axes
\begin{equation}
  \textrm{Kinetic Energy} =  \frac{3}{2} \textrm{Tr}[\dot{R}^T M \dot{R}]  = \frac{1}{2} \omega^T \mathcal{I}  \omega,
\end{equation}
where $M_{ij} \equiv \int d^3 x  \, \rho(x) \, x_i x_j$ and $\mathcal{I} \equiv 3 \textrm{Tr}[M] \mathds{1}  - M$.  Reflecting the fact that some angular directions are harder to move in, the metric becomes stretched in those directions 
\begin{equation}
ds^2 = 3 \textrm{Tr}[ dR^{T}  M  dR ] . \label{eq:metricrigidrotation}
\end{equation}
The free motion of a rigid body traces out the geodesics of this squashed three-sphere. (Specifically, the polhode rolls without slipping on the herpolhode lying in the invariable plane.)

\subsubsection{SO(3) = SU(2)/$\mathbb{Z}_2$}
The metrics Eq.~\ref{squashedunitarysphere} and Eq.~\ref{eq:metricrigidrotation} are (almost) the same. Underlying the link between the geometry of unitaries acting on a single qubit and the geometry of rotations in three dimensions is the mathematical fact that SU(2) is the double cover of SO(3),
\begin{eqnarray}
\textrm{SU}(2) &=& \mathbb{S}^3 \\
 \textrm{SO}(3)&  =& \textrm{symmetries of }\mathbb{S}^2 \ = \ \mathbb{S}^3/\mathbb{Z}_2 .
\end{eqnarray}
The transformations of a single qubit may thus be mapped to the configurations of a rigid body, with the mapping given by
\begin{eqnarray}
R_{ij} & =&  \textrm{Tr}[U^\dagger \sigma_i U \sigma_j]  \label{eq:RintermsofU} \\
\omega_i & =& i  \textrm{Tr} [ \dot{U} U^\dagger \sigma_i ].
\end{eqnarray}
Both $U$ and $-U$ get mapped to the same element $R_{ij}$, reflecting the fact that SU(2) is the double cover of SO(3).
(Indeed, since $U$ and $-U$ differ only by a global phase, we might be tempted to change the definition of complexity geometry so as to identify them, moving from SU(2) to $\textrm{PU(2)} \equiv \textrm{U}(2)/\textrm{U}(1) = \textrm{SU}(2)/\mathbb{Z}_2 = \textrm{SO}(3)$, in which case the connection between complexity geometry and rigid motion becomes exact.)

 \begin{center}
\begin{tabular}{c||c||c} 
UNITARY COMPLEXITY  \  & \   RIGID BODY MOTION  \ & \ GEODESIC MOTION \\
of single qubit  \ & \  in three dimensions \ & \ on squashed three-sphere \Bo \\
\hline
\hline
space of single-qubit  \ & \ space of orientations   \ & \  space of points  \To\\
operators is SU(2) $=S^3$  \ & \ is SO(3) = SU(2)/$\mathbb{Z}_2$ $=S^3/\mathbb{Z}_2$ \ & is topologically $S^3$
   \Bo \\
\hline
some tangent directions are \  & \ some axes have  \ & \ some directions have \To  \\ 
easy/hard to move in \ &  \   small/large moment of inertia \Bo \ & \ small/large circumferences  \\
\hline
least complex way    \  & \ free rotation  \ & \ geodesic motion \To  \\ 
to compile unitary \ &  \   of rigid body \Bo \ & \ on squashed sphere  \\
\hline
if all tangent directions   \ & \ if all axes   \ & \  if all circumferences   \To\\
are equally penalized  \ & \  have same moment of inertia   \ & \  are equally squashed  \\
recover inner-product metric   \ & \ recover round SO(3) metric \ &  \ recover round $S^3$ metric
   \Bo \\
\hline

\end{tabular}
\end{center}

Lab-frame quantities like $R_{ij}$ are neither left-invariant nor right-invariant, in the sense that they change under both $U \rightarrow U \, U_R$ and $U \rightarrow U_L \, U$. By contrast body-frame quantities like $\omega_i$ are right-invariant, since they don't care about the overall orientation of the rigid body.

For the complexity geometry the coefficients $\mathcal{I}_{ij}$ encapsulate the extent to which some tangent space directions are easier or harder to move in. For the rigid body, this gets mapped to different axes being easier or harder to rotate around because of unequal moments of inertia. (For the complexity geometry the only restriction is that the coefficients of $\mathcal{I}_{ij}$ be positive; for a rigid body with non-negative mass-density there is a triangle inequality $0 \leq \mathcal{I}_{zz} \leq \mathcal{I}_{xx} + \mathcal{I}_{yy}$, so that, for non-negative mass-density, $\mathcal{I}_{xx} = \mathcal{I}_{yy} =1$ \& $\mathcal{I}_{zz} > 2$ is an impossible Berger.)

\subsection{Complexity geometry in Euler coordinates} \label{sec:unitaryinEuler}
Let's put a coordinate system on the unitary group in order to give an explicit representation of the complexity metric. We will use (the Tait-Bryan version of) Euler angles. Any element of SU(2) may be written as $U= e^{i \sigma_z \theta_z} e^{i \sigma_y \theta_y } e^{i \sigma_x \theta_x }$,   and plugging this parametrization into Eq.~\ref{squashedsphere} gives an explicit form for the metric.

\subsubsection{$\mathcal{I}_{ij} = \delta_{ij} \rightarrow$ round $S^3$}
In Euler coordinates, the inner-product metric is 
\begin{equation}
\textrm{inner product: } \ ds^2 \biggl|_{\mathcal{I}_{ij} = \delta_{ij}}=  d \theta_x^{\, 2} + d \theta_y^{\, 2}  + d \theta_z^{\, 2} + 2 \sin 2 \theta_y \, d \theta_x d \theta_z . \label{eq:homogeneousandisotropicmetric}
\end{equation}
This is the round three-sphere, in unusual coordinates. All points are the same, and all directions are the same. 

\subsubsection{$\mathcal{I}_{xx} =\mathcal{I}_{yy} = 1 \rightarrow$ Berger Sphere}
We may break the symmetry $SU(2) \rightarrow U(1)$ by making one of the moments of inertia different from the other two.
This gives the metric of the so-called Berger sphere \cite{bergersphere}
\begin{equation}
ds^2 \biggl|_{\mathcal{I}_{xx} = \mathcal{I}_{yy} = 1 }=  \cos^2 2 \theta_y d \theta_x^{\, 2} +  d \theta_y^{\, 2}  + \mathcal{I}_{zz}  (d \theta_z + \sin 2 \theta_y d \theta_x)^2   . \label{eq:BergerMetric}
\end{equation}
The geometry of Eq.~\ref{eq:BergerMetric} is also sometimes called a `squashed three-sphere', but it is not squashed in the same sense that a beachball gets squashed when you sit on it. A squashed beachball (an oblate spheroid, if you will) is neither isotropic nor homogeneous---the equator bulges more than the poles. A squashed sphere, by contrast, is also not isotropic but \emph{is} homogeneous---though not all directions are the same, all points are.  We will have more to say about this---and how it relates to the right-but-not-left-invariance of Eq.~\ref{squashedsphere}---in Sec.~\ref{sec:leftandright}.\footnote{Beyond Berger: for general $\mathcal{I}_{ij}$ the components of the complexity metric in Euler coordinates are 
\begin{equation}
g_{ij} = \left( \begin{array}{ccc}
g_{xx}  & (\mathcal{I}_{xx} - \mathcal{I}_{yy})  \cos 2 \theta_y \cos 2 \theta_z \sin 2 \theta_z  &  \mathcal{I}_{zz} \sin 2 \theta_y \\
(\mathcal{I}_{xx} - \mathcal{I}_{yy})  \cos 2 \theta_y \cos 2 \theta_z \sin 2 \theta_z &  \mathcal{I}_{xx} \sin^2 2 \theta_z  + \mathcal{I}_{yy} \cos^2 2 \theta_z & 0  \\
\mathcal{I}_{zz} \sin 2 \theta_y   & 0 & \mathcal{I}_{zz} 
\end{array} \right) ,
\end{equation} 
where $g_{xx} = \cos^2 2 \theta_y( \mathcal{I}_{xx}  \cos^2 2 \theta_z + \mathcal{I}_{yy} \sin^2 2 \theta_z)  + \mathcal{I}_{zz} \sin^2 2 \theta_y$. 
This is the fully general squashed three-sphere. Though it is again not obvious from inspection, this metric is  completely homogeneous but also completely anisotropic. 
The sectional curvature is minus the deviation of a vector in the $i${th} direction due to displacement in the $j$th direction, namely \cite{Rajeev}
\begin{equation}
\textrm{geodesic deviation}_{ij} = - \mathcal{R}_{i j}^{\ \ j i} \biggl|_\textrm{not summed}= \frac{3 \mathcal{I}_{kk}^2 - 2 (\mathcal{I}_{ii} + \mathcal{I}_{jj}) \mathcal{I}_{kk}- (\mathcal{I}_{ii} - \mathcal{I}_{jj})^2  }{ \mathcal{I}_{ii} \mathcal{I}_{jj} \mathcal{I}_{kk}} \biggl|_\textrm{not summed}. \label{eq:geodesicdeviationSO3}
\end{equation}}

For the inner-product metric of Eq.~\ref{eq:homogeneousandisotropicmetric}, all the sectional curvatures are positive---indeed we know that the inner-product metric gives a uniformly positively curved three-sphere---which means that geodesics converge. However, for the Berger sphere with large enough $\mathcal{I}_{zz}$, sections defined by two easy directions become negatively curved
\begin{equation}
\hspace{-4cm} \textrm{at } \theta_x = \theta_y = \theta_z=0: \ \ \  \ \ \ \ \ \ \mathcal{R}_{\theta_x \theta_y}^{\ \ \ \theta_y \theta_x} = \mathcal{R}_{\theta_y \theta_x}^{\  \ \ \theta_x \theta_y}  = 4 - 3 \mathcal{I}_{zz},
\end{equation}
which means that easy geodesics diverge. We will discuss the significance of this in Sec.~\ref{sec:whynegativecurvature}. 

(Even as $\mathcal{I}_{zz}$ gets large, the curvature of sections defined by one easy and one hard direction remains positive, $\mathcal{R}_{\theta_x \theta_z}^{\ \ \ \theta_z \theta_x} = \mathcal{R}_{\theta_y \theta_z}^{\ \ \ \theta_z \theta_y} = \mathcal{I}_{zz}$, but not positive enough to prevent the Ricci scalar $\mathcal{R} =  \mathcal{R}_{\theta_y \theta_x}^{\  \ \ \theta_x \theta_y}  +  \mathcal{R}_{\theta_x \theta_y}^{\  \ \ \theta_y \theta_x}  +  \mathcal{R}_{\theta_y \theta_z}^{\ \ \ \theta_z \theta_y} +  \mathcal{R}_{\theta_z \theta_y}^{\ \ \ \theta_y \theta_z}  +  \mathcal{R}_{\theta_x \theta_z}^{\ \ \ \theta_z \theta_x}  +  \mathcal{R}_{\theta_z \theta_x}^{\ \ \ \theta_x \theta_z}   = 8 - 2 \mathcal{I}_{zz}$ going negative.)

%

\section{State-Complexity Geometry of a Single Qubit} \label{sec:stateC}
\sc

In the last section we considered the computational complexity of \emph{unitaries}. The space of SU(2) unitaries is topologically an $\mathbb{S}^3$ (and, for the inner-product metric but not the complexity metric, also geometrically an $\mathbb{S}^3$). In this section, we will consider the computational complexity of \emph{pure states}. The $\mathbb{CP}^1$ space of one-qubit pure states is topologically an $\mathbb{S}^2$ (and, for the inner-product metric but not the complexity metric, also geometrically an $\mathbb{S}^2$, the Bloch sphere). 

The space of single-qubit pure states has one fewer dimension since any given unitary has a kernel---a state it leaves invariant. Thinking of the unitary as a rotation, this kernel is the vector that points down the rotation axis. Therefore a given infinitesimal step in Hilbert space may be effected by any member of a one-parameter family of unitaries---to calculate the complexity metric on state space, we will want to find the shortest unitary in this family. 

\subsection{Deriving the one-qubit state-complexity metric}

The state of the qubit will be represented with $\vec{\psi}$, a unit-vector in $\mathbb{R}^3$. In this representation the states in the Hilbert space form an $\mathbb{S}^2$: the Bloch sphere. The map from the $\{ | 1 \rangle, |0 \rangle \}$ representation of the quantum state to the Bloch sphere representation is given by 
\begin{equation}
\vec{\psi} = \langle \psi | \vec{\sigma} | \psi \rangle.
\end{equation}
The Hilbert space inner product of two states defined in this way is 
\begin{equation}
|\langle \psi_2 | \psi_1 \rangle|^2 
 = \frac{1+ \vec{\psi}_1 \cdot \vec{\psi}_2}{2} = \frac{1 + \cos  \theta_{12}}{2} = \cos^2 \frac{\theta_{12}}{2}. \label{eq:halfangle}
\end{equation}
To compute the metric it suffices to consider target states that are nearby, $|\psi + {d \psi} \rangle$;  normalization requires that $\vec{\psi} \cdot  \vec{d \psi} = 0$. To leading order, the one-parameter family that transforms from $|\psi \rangle$ to $|\psi + {d \psi} \rangle$ is 
\begin{equation}
U = \mathds{1} + \frac{1}{2} i (\vec{d \psi} \times \vec{\psi} + \alpha \vec{\psi}) \cdot \vec{\sigma}  + \ldots. \label{eq:unitaryfamilyalpha}
\end{equation}
Geometrically we can understand this unitary as rotating the state around an axis 
\begin{equation}
\vec{r} = \frac{ \hat{d \psi} \times \vec{\psi} + \alpha \vec{\psi}}{|\hat{d \psi} \times \vec{\psi} + \alpha \vec{\psi}|} \label{eq:rofalpha}
\end{equation}
that is orthogonal to $\vec{d \psi}$ (but not necessarily orthogonal to $\vec{\psi}$); see Fig.~\ref{fig-VariousAxes}. \\

\noindent With no penalty factor ($\mathcal{I}_{ij} = \delta_{ij}$), the metric is 
\begin{equation}
ds^2 =  \textrm{Tr}[ dU^\dagger  dU ] = \frac{(\vec{d \psi} \times \vec{\psi} + \alpha \vec{\psi}) \cdot (\vec{d \psi} \times \vec{\psi} + \alpha \vec{\psi})}{4} = \frac{\vec{d \psi} \cdot \vec{d \psi} + \alpha^2}{4}.
\end{equation}
This is minimized at $\alpha = 0$, giving the inner-product metric $4ds^2 =  \vec{d \psi} \cdot \vec{d \psi} = d \theta^2 + \sin^2 \theta d \phi^2$. (The factor of 4 in this expression is because the distance is \emph{half} the angle, as in Eq.~\ref{eq:halfangle}, reflecting that orthogonal states, $\Delta s = \pi/2$, appear on  the Bloch sphere at antipodal points, $\Delta \psi = \pi$.)\\

\noindent Now let's introduce a penalty factor, $\mathcal{I}_{ij} = \delta_{ij} + ( \mathcal{I}_{zz} - 1) {p}_i {p}_j$, that punishes rotations around the axis $\vec{p}$ while leaving rotations round axes orthogonal to $\vec{p}$ unpunished. The penalized scalar product  may be defined as $\vec{a} \diamond \vec{b} \equiv a_i \mathcal{I}_{ij} b_j = \vec{a} \cdot \vec{b} + (\mathcal{I}_{zz} - 1) (\vec{a} \cdot \vec{p})(\vec{b} \cdot \vec{p})$. 
Equation~\ref{squashedunitarysphere} gives the complexity $ds$ of implementing $dU$ as  
\begin{eqnarray}
ds^2 &=&   \textrm{Tr}[i dU U^\dagger  \sigma_i] \mathcal{I}_{ij} \textrm{Tr}[i dU U^\dagger  \sigma_j] \\
4 ds^2 &=&  {(\vec{d \psi} \times \vec{\psi} + \alpha \vec{\psi}) \diamond (\vec{d \psi} \times \vec{\psi} + \alpha \vec{\psi})}\\ 
4 ds^2  & = & \vec{d \psi} \times \vec{\psi}  \diamond \vec{d \psi} \times \vec{\psi}  - \frac{(\vec{\psi} \diamond \vec{d \psi} \times \vec{\psi} )^2}{\vec{\psi} \diamond \vec{ \psi}} + \vec{\psi} \diamond \vec{\psi} \left(\alpha + \frac{\vec{\psi} \diamond \vec{d \psi} \times \vec{\psi} }{ \vec{\psi} \diamond \vec{ \psi}} \right)^2. \label{eq:dsofalpha}
\end{eqnarray}

If we set $\alpha = 0$ by fiat, then the metric is $4 ds^2 = d\theta^2 + \mathcal{I}_{zz}\sin^2 \theta d \phi^2$: motion down a line of longitude is unpenalized; motion around a line of latitude is penalized by $\sqrt{ \mathcal{I}_{zz}}$; and there is a conical singularity at the origin. However, choosing $\alpha = 0$ is choosing to go the long way around, because we can generally shorten distances by judicious choice of $\alpha$.

To derive the state complexity geometry, Eq.~\ref{eq:stateintermsoffunction} tells us to choose the member of the one-parameter family of connecting unitaries that minimizes the complexity, which means choosing $\alpha$ to put the last term in Eq.~\ref{eq:dsofalpha} to zero. Then using polar coordinates  on the two-sphere that place the penalized axis $\vec{p}$ at the north pole gives the metric for $\mathcal{I}_{xx} = \mathcal{I}_{yy} = 1$ as\footnote{For general $\mathcal{I}_{ij}$ we should reuse Eq.~\ref{eq:dsofalpha} except now with $\vec{a} \diamond \vec{b} \equiv \mathcal{I}_{xx} (\vec{a} \cdot \vec{p}_x)(\vec{b} \cdot \vec{p}_x) + \mathcal{I}_{yy} (\vec{a} \cdot \vec{p}_y)(\vec{b} \cdot \vec{p}_y) + \mathcal{I}_{zz} (\vec{a} \cdot \vec{p}_z)(\vec{b} \cdot \vec{p}_z)$.
This gives the unilluminating expression
\begin{eqnarray}
4 ds^2 & = & \frac{\mathcal{I}_{zz} \mathcal{I}_{yy}  \cos^2 \theta \cos^2 \phi + \mathcal{I}_{zz} \mathcal{I}_{xx} \cos^2 \theta \sin^2 \phi + 
 \mathcal{I}_{xx} \mathcal{I}_{yy} \sin^2 \theta }{\mathcal{I}_{zz} \cos^2 \theta + \mathcal{I}_{xx} \sin^2 \theta \cos^2 \phi + \mathcal{I}_{yy} \sin^2 \theta \sin^2 \phi } d\theta^2 \nonumber \\
 &&  \ \ \ \ +   2 \frac{ \mathcal{I}_{zz} (\mathcal{I}_{xx} - \mathcal{I}_{yy}) \sin \phi \cos \phi  \sin \theta \cos \theta  }{\mathcal{I}_{zz} \cos^2 \theta + \mathcal{I}_{xx} \sin^2 \theta \cos^2 \phi + \mathcal{I}_{yy} \sin^2 \theta \sin^2 \phi } d\theta d \phi  \nonumber \\
&& \ \ \ \ \ \ \ \ + \frac{ \mathcal{I}_{zz} ( \mathcal{I}_{xx} \cos^2 \phi + \mathcal{I}_{yy} \sin^2 \phi) \sin^2 \theta  }{\mathcal{I}_{zz} \cos^2 \theta + \mathcal{I}_{xx} \sin^2 \theta \cos^2 \phi + \mathcal{I}_{yy} \sin^2 \theta \sin^2 \phi }d \phi^2  \ \ .
\end{eqnarray}}
\begin{equation}
4 ds^2 =    \vec{d \psi} \cdot \vec{d \psi} +  \frac{(\mathcal{I}_{zz} -1)(\vec{d \psi} \cdot \vec{\psi} \times \vec{p})^2}{ (\mathcal{I}_{zz}-1) (\vec{\psi} \cdot \vec{p})^2 +1 } = d \theta^2 + \frac{\mathcal{I}_{zz} \sin^2 \theta}{\mathcal{I}_{zz} \cos^2 \theta  + \sin^2 \theta} d \phi^2. \label{eq:statespacecomplexitymetric}
\end{equation}
For $\mathcal{I}_{zz}=1$, all directions are treated the same and we recover the standard round `Fubini-Study' metric on the Bloch sphere. For $\mathcal{I}_{zz} \neq 1$, however, the complexity metric differs from the inner-product metric. We will study the properties of this metric in Sec.~\ref{sec:propertiesofstatecomplexity}.

\subsubsection{Geometric interpretation}
\begin{figure}[htbp] 
   \centering
   \includegraphics[width=.85\textwidth]{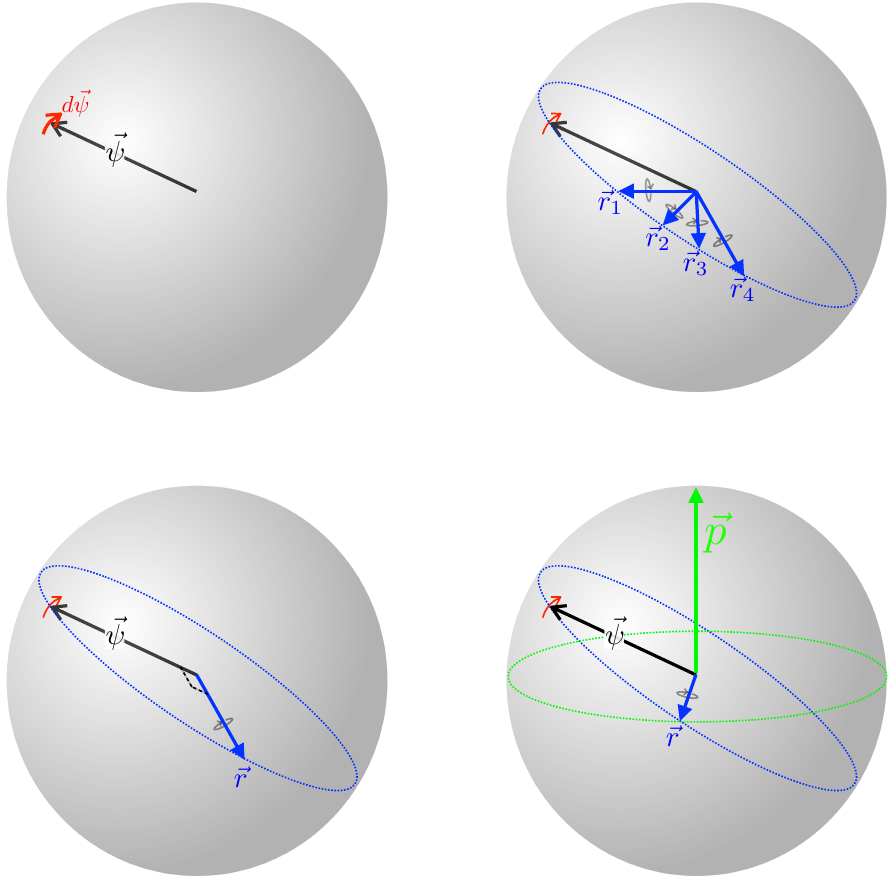} 
   \caption{Top left: the state of the qubit can be represented as a unit vector $\vec{\psi}$ on the two-sphere. 
Top right: to transform  from $|\vec{\psi} \rangle$ to $|\vec{\psi} +  {\color{red} \vec{ d \psi} }\rangle$ we can rotate around any axis ${\color{blue}  \vec{r}}$ that is orthogonal to the change in state, ${\color{blue}  \vec{r}} \cdot {\color{red}  \vec{ d \psi }} =0$. There is a one-parameter family of possible rotation axes that  all satisfy ${\color{blue}  \vec{r}} \cdot {\color{red}  \vec{ d \psi }} =0$ and therefore any of which can be used to implement the transformation; these lie on a great circle. 
   Bottom left: in order to make the rotation angle $d \theta$ as small as possible, we should rotate around an axis orthogonal to the initial state, $\vec{\psi} \cdot {\color{blue} \vec{r} } = 0$; this is the appropriate rotation axis when $\mathcal{I}_{zz} = 1$. Bottom right: in order to make the penalty factor as small as possible, we should rotate around an axis orthogonal to the penalized direction, ${\color{green} \vec{p}} \cdot {\color{blue} \vec{r} } = 0$; this is the appropriate rotation axis when $\mathcal{I}_{zz} \rightarrow \infty$, as in Sec.~\ref{subsec:infiniteIzz}. For intermediate values, $1<\mathcal{I}_{zz}<\infty$, the optimal rotation axis, given by Eq.~\ref{eq:optimalalphaaxis},  lies between these two extremes.}
   \label{fig-VariousAxes}
\end{figure}

To add insight, let's re-examine the derivation of Eq.~\ref{eq:statespacecomplexitymetric} while stressing its geometrical interpretation. The essential idea is captured by Fig.~\ref{fig-VariousAxes}.

To transform from $\vec{\psi}$ to $\vec{\psi} + \vec{d \psi}$ (where $\vec{d \psi} \cdot \vec{\psi} = 0$ by conservation of normalization) we must rotate around an axis $\vec{r}$ that is orthogonal to $\vec{d \psi}$
\begin{equation}
\vec{d \psi} \cdot \vec{r} = 0 \ . 
\end{equation}
Any such axis will work, and there is a one-parameter family of them, given by Eq.~\ref{eq:rofalpha}. 

Elementary geometry tells us that, to get from $\vec{\psi}$ to $\vec{\psi} + \vec{d \psi}$ by rotating around the axis $\vec{r}$, the total angle $d \theta$ through which we must rotate is given by 
\begin{equation}
d \theta^2 = \frac{\vec{d \psi} \cdot \vec{d \psi} }{\, |\vec{\psi} \times \vec{r} \, |^2} .
\end{equation}
This angle is minimized when the axis of rotation $\vec{r}$ is orthogonal to the initial state $\vec{\psi}$, so that $|\vec{\psi} \times \vec{r}| = 1$. This is equivalent to putting $\alpha = 0$. When there is no penalty factor this is the optimal rotation axis, and this gives the inner-product metric.

However, once $\mathcal{I}_{zz} \neq 1$ not all $d \theta$s count equally. Instead, the complexity distance $ds$ associated with a given $d \theta$ varies with the axis of rotation as 
\begin{equation}
4 ds^2 = \left( 1 + (\mathcal{I}_{zz} - 1) (\vec{p} \cdot \vec{r})^2 \right) d \theta^2 .
\end{equation}
For $\mathcal{I}_{zz} > 1$, the larger $|\vec{p} \cdot \vec{r}|$, the more the rotation is penalized.

As a function of $\vec{r}$, the complexity distance associated with a given $\vec{d \psi}$ is therefore 
\begin{eqnarray}
4 ds^2 &= & \frac{ 1 + (\mathcal{I}_{zz} - 1) (\vec{p} \cdot \vec{r})^2 }{|\vec{\psi} \times \vec{r} \, |^2} \vec{d \psi} \cdot \vec{d \psi}.
\end{eqnarray}
The optimal rotation axis is determined by the interplay of two factors. On the one hand, to make the rotation angle $d \theta$ small we should choose a rotation axis that is as close as possible to orthogonal to the initial state, $\vec{r} \cdot \vec{\psi} = 0$. On the other hand, to make the penalty factor small we should choose a rotation axis that is as close as possible to orthogonal to the penalized direction $\vec{r} \cdot \vec{p} = 0$. The optimal compromise is given by 
\begin{equation}
\alpha =   \frac{(\mathcal{I}_{zz} - 1)( \vec{p} \cdot \vec{d \psi} \times \vec{\psi} )( \vec{p} \cdot \vec{\psi})}{1 + (\mathcal{I}_{zz} - 1) \vec{p} \cdot \vec{\psi} }, \label{eq:optimalalphaaxis}
\end{equation}
which not-coincidentally is also the value of $\alpha$ that puts the last term in Eq.~\ref{eq:dsofalpha} to zero. When  $\mathcal{I}_{zz} = 1$ the optimal axis of rotation is orthogonal to the initial vector $\vec{r} \sim \vec{\psi} \times \hat{d \psi}$. When $\mathcal{I}_{zz} \gg 1$, the axis of rotation is forced to be equatorial, $\vec{r} \cdot \vec{p} = 0$. 
For any value of $\mathcal{I}_{zz}$, rotating around the optimal axis recovers the complexity metric of Eq.~\ref{eq:statespacecomplexitymetric}.

\FloatBarrier
\subsection{Properties of the one-qubit state-complexity metric} \label{sec:propertiesofstatecomplexity}

\begin{figure}[htbp] 
   \centering
   \includegraphics[width=5in]{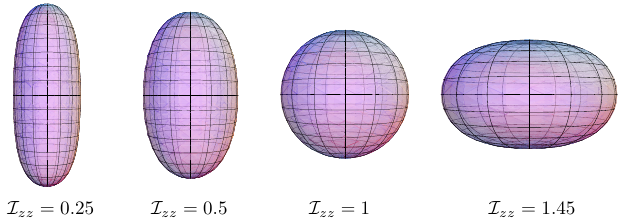} 
   \caption{Deformed Bloch spheres with various penalty factors $\mathcal{I}_{zz}$. Unlike the `squashed'  three-spheres of Sec.~\ref{sec:unitaryinEuler}, these two-spheres \emph{are} squashed in the same sense that a beachball gets squashed when you sit on it.  For $\mathcal{I}_{zz}=1$ every direction is punished equally and we have the Bloch sphere with the standard inner-product metric. The Bloch spheroid is prolate for $\mathcal{I}_{zz}<1$ and oblate for $\mathcal{I}_{zz}>1$. For $\mathcal{I}_{zz}>3/2$ the Bloch spheroid is negatively curved at the poles and cannot be embedded in $\mathbb{R}^3$. For very large $\mathcal{I}_{zz}$ the Bloch spheroid becomes two back-to-back negatively curved spaces glued together by positive curvature at the equator. }
   \label{fig-elongatedblochspheres}
\end{figure}

The complexity metric for a one-qubit state with $\mathcal{I}_{xx} = \mathcal{I}_{yy} = 1$ is described by Eq.~\ref{eq:statespacecomplexitymetric}. The effect of the penalty factor is to deform the Bloch sphere. As shown in Fig.~\ref{fig-elongatedblochspheres}, for $\mathcal{I}_{zz} < 1$ motion along lines of latitude becomes easier and the Bloch spheroid becomes prolate\footnote{This might be a good metric to describe historical trade and migration patterns on Earth, where there is evidence that climate gradients make East-West spreading easier than North-South spreading \cite{EastWest}.}; for $\mathcal{I}_{zz} > 1$ motion along lines of latitude becomes harder and the Bloch spheroid becomes oblate.

\begin{figure}[htbp] 
   \centering
   \includegraphics[width=.52\textwidth]{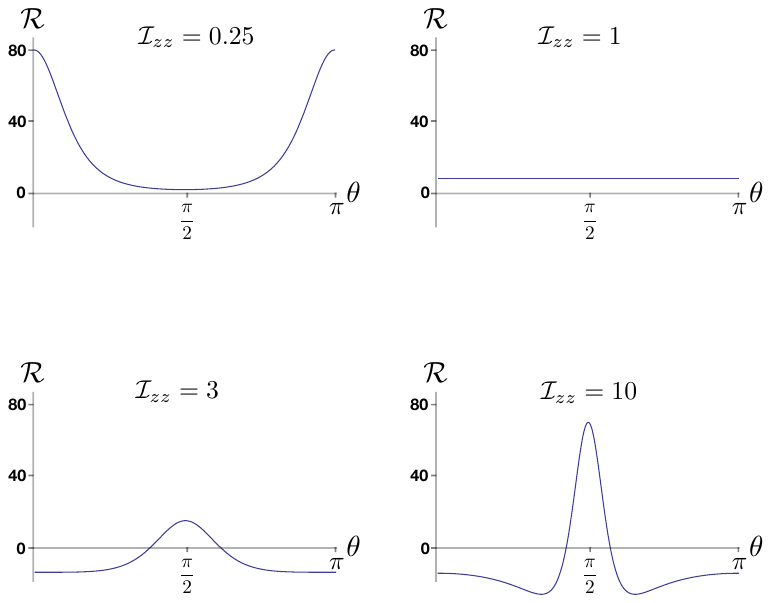} 
   \caption{The curvature $\mathcal{R}$ of the deformed Bloch sphere for different values of $\mathcal{I}_{zz}$, given by Eq.~\ref{eq:curvatureofBloch}. For $\mathcal{I}_{zz} < 1$ the curvature is everywhere positive and largest at the poles.  For $\mathcal{I}_{zz}=1$ the curvature is uniformly positive and the Bloch sphere is spherical. For large $\mathcal{I}_{zz}$, the curvature is like two negatively curved disks stuck together by a thin very positively curved wall.}
   \label{fig-complexityofstateplot}
\end{figure}

The curvature of this metric is 
\begin{eqnarray}
\mathcal{R} &=& \frac{8 \mathcal{I}_{zz}( 1 - 2 (\mathcal{I}_{zz}-1)\cos^2 \theta) }{(\mathcal{I}_{zz} \cos^2 \theta + \sin^2 \theta)^2}  \label{eq:curvatureofBloch}\\
\textrm{pole}: \ \mathcal{R} \biggl|_{\theta=0} \, \hspace{1pt}& =& 8 \left( \frac{3}{\mathcal{I}_{zz}} - 2 \right)  \\
\textrm{equator}: \ \mathcal{R} \biggl|_{\theta=\frac{\pi}{2}} &=& 8 \mathcal{I}_{zz}.
\end{eqnarray}
For $\mathcal{I}_{zz}<\frac{3}{2}$ the curvature is everywhere positive. But, as is shown in Fig.~\ref{fig-complexityofstateplot}, for large  $\mathcal{I}_{zz}$ the curvature starts out negative at the pole, becomes {more} negative away from the pole, before hitting a minimum just above the equator at
\begin{equation}
\partial_{\theta} \mathcal{R} = 0 \ \rightarrow \ \cos^2 \theta_\textrm{min} = \frac{2}{\mathcal{I}_{zz}-1} \ \  \rightarrow \ \  
 \mathcal{R}_\textrm{min} = - \frac{8\mathcal{I}_{zz}}{3}.
\end{equation}
The curvature then shoots up to a large positive value $8 \mathcal{I}_{zz}$ at the equator. 

\FloatBarrier

\subsection{Even simpler case: one-qubit complexity with $\mathcal{I}_{zz} \rightarrow \infty$} \label{subsec:infiniteIzz}
Let us consider the state complexity of a single qubit in the limit in which $\sigma_z$-rotations are completely forbidden, $\mathcal{I}_{zz} \rightarrow \infty$: the $z$-axis of the gimbal has completely seized up and will not rotate at all, and so the rotation axis $\vec{r}$ is required to be equatorial. 

Everywhere \emph{except} very close to the equator, the state complexity geometry becomes 
\begin{eqnarray}
\mathcal{I}_{xx} = \mathcal{I}_{yy} = 1 \ \& \ \mathcal{I}_{zz} \rightarrow \infty: \ \ \ \ \ \ \ \ \ \ 4 ds^2 &=& d \theta^2 + \tan^2 \theta d \phi^2. \ \ \ \ \ \   \ \ \ \ \ \ \ \ \ \ \ \ \ \ \ \ \   \label{eq:statespaceIinfinity} \\
\rightarrow \ \  \mathcal{R} &=& - \frac{16}{\cos^2 \theta}.
\end{eqnarray}
The one-qubit state complexity geometry is almost-everywhere negatively curved as $\mathcal{I}_{zz} \rightarrow \infty$. However, nowhere does the geometry closely resemble the hyperbolic plane: there are no regions many-curvature-lengths long where the curvature length ($\ell = \cos \theta$) is roughly constant.

Very close to the equator, $\theta \sim \frac{\pi}{2} \pm O(\frac{1}{\sqrt{\mathcal{I}_{zz}}}$), the metric deviates from Eq.~\ref{eq:statespaceIinfinity} and has a delta-function positive curvature. For the unitary complexity metric of Sec.~\ref{sec:unitaryinEuler}, when $\mathcal{I}_{zz} \rightarrow \infty$ the curvature diverged everywhere; for the state complexity it only diverges at the equator. In the limit $\mathcal{I}_{zz} \rightarrow \infty$, the complexity state geometry becomes two negatively curved disks  glued together along the equator. 

As with the unitary complexity of the last section, we have seen that non-trivial $\mathcal{I}_{ij}$ can deform a positively curved sphere such that sectional curvatures become negative. We will return to this again in Sec.~\ref{sec:whynegativecurvature}.

The separation of two points both on the equator  is
\begin{equation}
\Delta s \Bigl|_{\theta_1 = \theta_2 = \frac{\pi}{2} , \, \mathcal{I}_{zz} = \infty, \, \mathcal{I}_{xx} = \mathcal{I}_{yy} = 1} = \frac{1}{2} \sqrt{ \Delta \phi (2 \pi - \Delta \phi)}, \label{eq:distancesimplestcaseequator}
\end{equation}
which is larger than $\Delta s \bigl|_{\theta_1 = \theta_2 = \frac{\pi}{2} , \, \mathcal{I}_{zz} = \mathcal{I}_{xx} = \mathcal{I}_{yy} = 1} = \frac{1}{2} \Delta \phi$ except at $\Delta \phi = 0$ and $\Delta \phi = \pi$.

\FloatBarrier
\section{Discussion} \label{sec:discussion}
\sc

The single-qubit example is rich enough to exhibit many of the essential features of complexity geometry, as we will now assess.

\subsection{Right-invariance \emph{vs}. left-invariance}  \label{sec:leftandright}

\noindent The one-qubit example accurately captures that the \emph{unitary} complexity distance is right-invariant but not left-invariant, and that the unitary complexity metric is homogeneous; the same example also captures that the \emph{state} complexity distance is not left-invariant, and that therefore the corresponding metric is not homogeneous. These properties apply both to complexity geometry and to the gate  definition of complexity. Table~\ref{Table1} gives a summary.

\subsubsection{Unitary complexity geometry on SU($2^N$) is homogeneous but anisotropic}

The \emph{inner-product} distance on SU($2^N$) is both left- and right-invariant since it is invariant under $\{ U_1, U_2 \} \rightarrow \{ U_L U_1 U_R, U_L U_2 U_R \}$, 
\begin{equation}
\textrm{left- and right-invariant: } \ \   \textrm{Tr}[(U_L U_1 U_R) (U_L U_2 U_R)^\dagger]  = \textrm{Tr}[U_1 U_2^{\dagger}]. \ \ \ \ \ \ 
\end{equation}
The right-invariance makes the inner-product metric on SU($2^N$) \emph{homogeneous} (i.e.~every point is the same). For the single-qubit example of Eq.~\ref{eq:homogeneousandisotropicmetric}, the inner-product metric on SU(2) is that of the round $S^3$, and so the SU(2) metric is in addition \emph{isotropic} (i.e.~every direction is the same); for $N  \geq  2$, the inner-product metric on SU($2^N$) is not isotropic.\\

\noindent The \emph{complexity} distance on SU($2^N$) is right-invariant but not left-invariant. 
The complexity distance between two unitaries $U_1$ and $U_2$ is given by the complexity of the unitary that connects them
\begin{equation}
\mathcal{C}[U_1; U_2]  = \mathcal{C}[U_{12}] \textrm{ where } U_1 = U_{12} U_2.
\end{equation}
(Taking $U_1$ and $U_2$ to be infinitesimally separated gives the complexity metric.) It is clear that $U_1 = U_{12} U_2$ if and only if $U_1 U_R = U_{12} U_2 U_R$, which means that the complexity is right-invariant:
\begin{equation}
\textrm{right-invariant: }  \ \ \mathcal{C}[U_1; U_2]  = \mathcal{C}[U_1 U_R; U_2 U_R] .
\end{equation}
We see this reflected in Eq.~\ref{squashedunitarysphere} where taking $U \rightarrow U U_R$ and $dU \rightarrow dU U_R$ does not change the complexity distance. By contrast, $U_1 = U_{12} U_2$ does not imply $U_L U_1 = U_{12} U_L U_2$, so in general the complexity is not left-invariant:
\begin{equation}
\textrm{not left-invariant: } \ \  \mathcal{C}[U_1; U_2]  \neq \mathcal{C}[U_L U_1 ; U_L U_2 ] .
\end{equation}
The asymmetry between left and right arises because $U_{12}$ acts from the left and not from the right---we compile circuits by adding gates to the end, not to the start. (Beware that in much of the mathematics literature this convention is reversed \cite{Milnor}.) Right-invariance of the complexity distance translates into \emph{homogeneity} of the complexity geometry, since right-multiplying by $U_R$ can translate any point to any other. 

\subsubsection{State complexity geometry on $\mathbb{CP}^{2^N-1}$ is neither homogeneous nor isotropic}
The \emph{inner-product} distance on $\mathbb{CP}^{2^N-1}$ is  left-invariant since it is invariant under $\{ | \psi_1 \rangle, | \psi_2 \rangle \} \rightarrow \{ U_L | \psi_1 \rangle, U_L | \psi_2 \rangle \}$, 
\begin{equation}
\textrm{left-invariant: } \ \   (U_L | \psi_1 \rangle)^\dagger (U_L | \psi_2 \rangle)  = ( | \psi_1 \rangle)^\dagger ( | \psi_2 \rangle). \ \ \ \ \ \ 
\end{equation}
This makes the inner-product metric (also known as the `Fubini-Study' metric) on $\mathbb{CP}^{2^N-1}$ both \emph{homogeneous} (every point is the same) and  \emph{isotropic} (every direction is the same). Indeed, for the single-qubit example of Eq.~\ref{eq:statespacecomplexitymetric}, the inner-product metric on $\mathbb{CP}^1$ is that of the round $S^2$. \\

\noindent The \emph{complexity} distance on $\mathbb{CP}^{2^N-1}$ is not left-invariant, for the same reason that the unitary complexity on SU($2^N$) is not left-invariant. As a consequence the state complexity geometry Eq.~\ref{eq:statespacecomplexitymetric} is neither isotropic nor homogeneous.


 \begin{center}
 \begin{table}
\begin{tabular}{c||c|c||c|c} 
  & right-invariant & left-invariant & homogeneous & \ \  isotropic  \Bo \\
  \hline
  \hline
  inner-product metric  SU($2$)    &  \ding{51} & \ding{51} & \ding{51}  & \ding{51} \To    \\
inner-product metric SU($2^N$)    &  \ding{51} & \ding{51} & \ding{51}  & \ding{55}  \Bo   \\
    \hline
    complexity metric SU($2$)    & \ding{51} & \ding{55}  & \ding{51}  &  \ding{55} \To   \\
        complexity metric SU($2^N$)    & \ding{51} & \ding{55}  & \ding{51}  &  \ding{55}  \Bo  \\
\hline
\hline
inner-product metric  $\mathbb{CP}^1$    & {N/A}  & \ding{51}  & \ding{51}  & \ding{51}  \To    \\
inner-product metric  $\mathbb{CP}^{2^N-1}$    & {N/A}  & \ding{51}  & \ding{51}  & \ding{51}   \Bo   \\
    \hline
    complexity metric $\mathbb{CP}^1$    & N/A & \ding{55} & \ding{55} & \ding{55} \To   \\
        complexity metric $\mathbb{CP}^{2^N-1}$    & N/A & \ding{55} & \ding{55} & \ding{55}  \Bo  \\
\end{tabular}
\caption{Inner-product metrics are left-invariant, whereas complexity metrics are not.} \label{Table1}
\end{table}
\end{center}
\FloatBarrier
The complexity metric is not left-invariant. Instead, there is a preferred basis in which the penalty tensor $\mathcal{I}_{IJ}$ is diagonal. This should not surprise us. Nature itself also picks out a preferred basis, since the Hamiltonian of the Standard Model is local in some bases but not in others. Ultimately, it is this special structure to the interactions of fundamental physics that is inherited by our laboratory equipment and begets the preferred basis for complexity.

\subsection{Geodesics and time-independent Hamiltonians} \label{subsec:GeodesicsAndTimeIndependent}

The one-qubit example accurately captures the general relationship between the paths generated by time-independent Hamiltonians, the geodesics of the complexity geometry on SU$(2^N)$, and the geodesics of the complexity geometry on $\mathbb{CP}^{2^N-1}$.

\subsubsection{Geodesics of the unitary group \emph{vs}. time-independent Hamiltonians}

\begin{itemize}
\item  For the \emph{inner-product} metric on {unitaries}, every geodesic is generated by a time-independent Hamiltonian, and every path generated by a time-independent Hamiltonian is a geodesic of the inner-product metric. The relationship is one-to-one.

\item For the \emph{complexity} metric on {unitaries}, the relationship is much looser: to generate geodesics generally requires the use of a time-dependent Hamiltonian, and a time-independent Hamiltonian generally does not generate a geodesic. This can be seen by examining the equation of motion. For a geodesic of the complexity geometry, the rate of change of the Hamiltonian in the direction of any operator $K$ is \cite{Nielsen2}    
\begin{equation}
\langle \dot{H},K \rangle = i \langle H, [H,K] \rangle, \label{eq:equationofmotioningenerality}
\end{equation}
where $\langle A,B \rangle \equiv \sum_{I J} \textrm{Tr}[A \sigma_I] \mathcal{I}_{IJ} \textrm{Tr}[B \sigma_J]$. For SU(2) this reduces to the Euler equations for a spinning top, and the statement that complexity geodesics do not conserve $H$ is equivalent to the statement that spinning tops do not conserve  angular velocity $\vec{\omega}$.

However, there is a class of time-independent Hamiltonians that do generate geodesics: those Hamiltonians for which every nonzero term has the same penalty factor. In this case, the cyclic property of the trace guarantees that the right-hand side of Eq.~\ref{eq:equationofmotioningenerality} is zero. Rotating bodies that start spinning around a principal axis continue to spin around a principal axis. 
\end{itemize}

 To use an automotive simile, minimizing the unitary complexity is like optimizing the driving route. To minimize travel time, the optimal route must strike a balance between keeping the number of miles small and keeping the traffic speed high. Similarly, to minimize complexity distance, the geodesic must strike a balance between keeping the inner-product distance small and keeping the penalty factor low. Using a time-independent Hamiltonian minimizes the inner-product distance, which is only optimal if the penalty factors are the same. When traffic speeds differ, the geodesic gives the route that minimizes not driving miles but driving minutes.

\subsubsection{Geodesics on Hilbert space \emph{vs}. time-independent Hamiltonians} \label{sec:geoHilbertvsH}

\begin{itemize}
\item For the \emph{inner-product} metric on {states}, every geodesic is generated by a time-independent Hamiltonian, but most paths generated by time-independent Hamiltonians are \emph{not} geodesics of the inner-product metric. 

For example, the time-independent Hamiltonian $\sigma_z$ rotates about the poles of the Bloch sphere and so generates all the lines of latitude, but only one of those lines of latitude---the equator---is an actual geodesic. By contrast all geodesics (all great circles) may be generated by rotating around some axis.

\item For the \emph{complexity} metric on {states}, in general there is no connection between geodesics and time-independent Hamiltonians. For the example of Eq.~\ref{eq:statespacecomplexitymetric}, the only geodesics that are generated by time-independent Hamiltonians are either lines of longitude (generated by purely easy Hamiltonians) or the equator (generated by purely hard Hamiltonians).

Consider the geodesic that connects two states both on the equator ($\theta = \frac{\pi}{2}$) of the $\mathcal{I}_{zz} = \infty$ state-complexity metric of Sec.~\ref{subsec:infiniteIzz}. The infinite penalty factor constrains the instantaneous rotation axis to be equatorial (constrains the Hamiltonian to have no $\sigma_z$ component), and subject to this constraint the geodesic will be the connecting path  that rotates through the smallest total angle. First consider the path that has a fixed rotation axis (a time-independent Hamiltonian). This path is given by rotating around the axis that bisects the two states---the total rotation angle along this path is $\pi$, so the complexity distance is $\Delta s = \pi/2$ (independent of $\Delta \phi$). However, this is not the shortest path, and not a geodesic. Instead, along the actual geodesic the relative sizes of the $\sigma_x$ and $\sigma_y$ terms change, so the instantaneous rotation axis precesses along the equator. The geodesic is thus generated by a time-dependent Hamiltonian and has length given by Eq.~\ref{eq:distancesimplestcaseequator}. 


\end{itemize} 

\subsubsection{Geodesics of the unitary group \emph{vs}. geodesics on Hilbert space}

Equation~\ref{eq:stateintermsoffunction} tells us that for every right-invariant metric on SU$(2^N)$ there is a corresponding metric on $\mathbb{CP}^{2^N-1}$. How are the geodesics of the two spaces related? 

The relationship between the two sets of geodesics is the same for the complexity metric as it was for the inner-product metric, and the same for the many-qubit metric as it was for the single-qubit metric. 
On the one hand every geodesic of the Hilbert space  can be implemented by acting with a geodesic of the unitary group: for every geodesic $|\psi_g(t) \rangle$ there is a geodesic U$_g(t)$ such that $|\psi_g(t) \rangle = \textrm{U}_g(t) |\psi(0) \rangle$. On the other hand, implementing a geodesic U$_g(t)$ on a typical state does not generally give a geodesic of the state metric: $|\psi(t) \rangle = \textrm{U}_g(t) |\psi(0) \rangle$  is usually not  a geodesic.

Both of these facts can be understood from the definition of state complexity in Eq.~\ref{eq:stateintermsoffunction}. The relative state complexity of two
states is the complexity of the least complex unitary that maps one to the other. There are two minimizations implicit in this definition: for a given unitary, minimize over all the paths that give rise to that unitary; and then minimize over all unitaries that implement the desired transformation on the state. It is the first minimization that guarantees geodesics of the state geometry can be implemented by geodesics of the unitary geometry; it is neglecting the second minimization that means that applying a geodesic of the unitary metric to a state will generally not give a geodesic of the state metric.

\subsection{Negative curvature} \label{sec:whynegativecurvature}

The one-qubit example accurately captures that introducing  anisotropic penalty factors may generate negative curvature. \\

 The inner-product metric for a single qubit is spherical, so the curvature is  positive. The mean curvature is positive, and every sectional curvature is positive. 
However, we saw with Eqs.~\ref{eq:geodesicdeviationSO3} and \ref{eq:curvatureofBloch} that for sufficiently anisotropic penalty factors some sectional curvatures become negative.

For example, consider the state complexity metric of Eq.~\ref{eq:statespacecomplexitymetric}. Expanding near the North pole at $\theta = 0$ (i.e.~expanding near the state that has $\langle \sigma_z \rangle = 1$) gives
\begin{equation}
4 ds^2  = d \theta^2 + \left( \theta^2  + \left( \frac{2}{3} - \frac{1}{\mathcal{I}_{zz}} \right)  \theta^4 + \ldots \right) d \phi^2 .
\end{equation}
For small $\mathcal{I}_{zz}$ the coefficient of $\theta^4$ is negative, and so the curvature is positive. But for $\mathcal{I}_{zz}  > 3/2$ the sign flips. Two `easy' geodesics that start together at $\theta=0$ and emanate along lines of constant $\phi$ have a separation that grows like
\begin{equation}
2 \Delta s = \left(  \theta + \frac{1}{2} \left( \frac{2}{3} - \frac{1}{\mathcal{I}_{zz}} \right) \theta^3  + \ldots \right) \Delta \phi + \ldots.
\end{equation} 
For $\mathcal{I}_{zz}  > 3/2$ the geodesics accelerate apart---negative curvature makes geodesics diverge. 

Anisotropic penalty factors do not make \emph{all} sections negatively curved. Sections are defined by a pair of directions, and we saw in Eq.~\ref{eq:geodesicdeviationSO3} that when the pair is comprised of one easy and one hard direction, or two hard directions, the sectional curvatures are still positive. But when the section is defined by two easy directions, the sectional curvature becomes negative. 

The mathematical origin of the negative curvature is that the commutator of two easy directions is itself hard. To see this, consider 
the Baker-Campbell-Hausdorff formula, which tells us that, to second order in $t$,
\begin{equation}
\exp [ {iH_1 t}] \exp[iH_2 t  ]    = \exp \left[ i (H_1 + H_2)t -  \frac{1}{2} [H_1,H_2 ] t^2 + \ldots \right] . \label{eq:BCH}
\end{equation}
This formula can be used to relate the length of the hypotenuse of a right-angled triangle to the length of its two sides. If one triangle arm points in the $(-H_1)$-direction, and the other in the $H_2$-direction, then Eq.~\ref{eq:BCH} tells us that the hypotenuse points not only in the $H_1 + H_2$-direction but also has a component in the $[H_1,H_2]$-direction. If the commutator $[H_1,H_2]$ is hard, then the effect of the penalty factor  will be to make the hypotenuse longer.

\begin{figure}[htbp] 
   \centering
      \includegraphics[width=.5\textwidth]{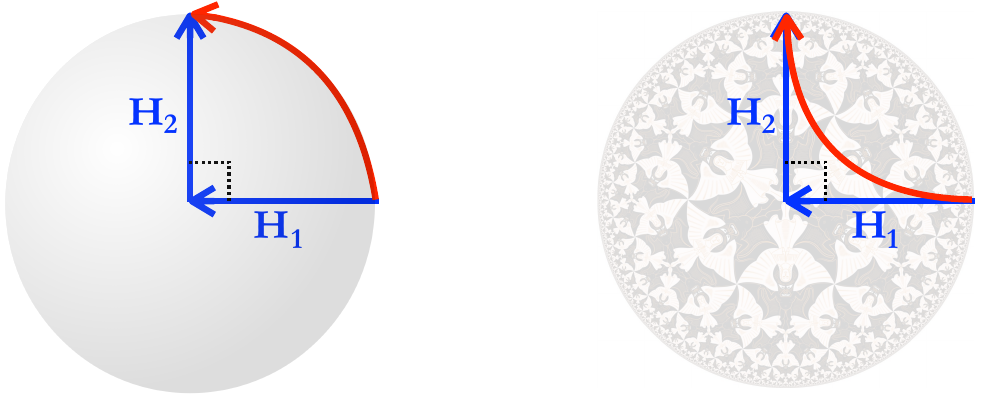} 
   \caption{Left: in a positively curved (or flat) space, taking the hypotenuse substantially shortens the path compared with following each arm in turn, ${\color{red} L_\textrm{hypot.}}  \leq {\color{blue} \sqrt{2} L_\textrm{arm}}$. Right: in a negatively curved space, the hypotenuse tends to hug each arm in turn---the more negative the curvature, the smaller the distance saved by taking the red path rather than the blue path, ${\color{red} L_\textrm{hypot.}}  \geq {\color{blue} \sqrt{2} L_\textrm{arm}}$.}
   \label{fig-hypotenuses}
\end{figure}

Long hypotenuses are a feature of negative curvature. Consider the right-angled triangles shown in Fig.~\ref{fig-hypotenuses}. In flat space, Pythagoras' theorem tells us that the hypotenuse (the red geodesic from one corner to the other) is much shorter than the sum of the two arms. In negatively curved spaces like the hyperbolic plane, by contrast, the hypotenuse tends to `hug' each of the two arms in turn, and has a length closer to the sum of the two arms and therefore longer than would have been predicted by Pythagoras.

In the complexity geometry, the hypotenuse is not much shorter than the sum of the two arms. In this regard, the complexity geometry reflects our expectations from  complexity. One way to compile $e^{i H_1 t} e^{i H_2 t}$ is to first compile $e^{i H_1 t}$ and then compile $e^{i H_2 t}$ and then concatenate the resulting unitaries---this is like taking the blue path in Fig.~\ref{fig-hypotenuses}. There may be a more efficient way to do it than simple concatenation---a red path---but generally the efficiencies are small, and this is reflected in the failure of the geodesic on the hypotenuse to substantially shorten the path. \\

Geodesics deviate on negatively curved spaces, and the deviation causes geodesic instability.  Rigid body motion also has a well-known instability---the ``tennis-racket effect''---in which rotation around the principal axis with intermediate moment of inertia is unstable. Given that we have shown that evolution on the complexity geometry is isomorphic to rigid body motion, you might wonder if these two instabilities are the same. They are not. For example, the tennis-racket instability is absent when $\mathcal{I}_{xx} = \mathcal{I}_{yy}$, whereas the negative curvature persists. \\

We have seen that the easy-easy section of the single-qubit complexity geometry is negatively curved. In the full $N$-qubit complexity geometry described in Sec.~\ref{sec:complexitygeometry}, it is also true that sections defined by two easy (low-$k$) directions are generally negatively curved. For the $N$-qubit complexity geometry the negative curvature gives rise to the chaotic growth of perturbations and is necessary to recover the `switchback' effect.  The connection between complexity and negative curvature is explored in more detail in Ref.~\cite{Brown:2016wib}.

\subsection{Cut loci and `tacking'} \label{sec:cutloci}

Another feature of complexity geometry that the one-qubit example manifests is the existence and importance of cut loci. (This section is based on material that was explained to us by Henry Lin.)

Consider the complexity of the unitary $U = e^{i \sigma_z \delta }$, for $\delta \ll 1$. One way to generate this unitary is straightforward: rotate directly in the $\sigma_z$-direction through an angle $\delta$. The length of this path upper-bounds the complexity distance 
\begin{equation}
\mathcal{C}[e^{i \sigma_z \delta} ] \leq \sqrt{\mathcal{I}_{zz}}  \delta. \label{eq:hardroutelocus}
\end{equation}

This is the most direct way to generate $U$. But it is not the only way, and for large $\mathcal{I}_{zz}$ may not be the cheapest way. An alternative path is to zigzag in the easy directions, commutating $\sigma_x$ and $\sigma_y$ to build up $\sigma_z$ indirectly. To leading order in $\delta$ we have 
\begin{equation}
e^{-i \sigma_x \sqrt{\delta/2}}e^{-i \sigma_y \sqrt{\delta/2}}e^{i \sigma_x \sqrt{\delta/2}}e^{i \sigma_y \sqrt{\delta/2}} = e^{ i \delta \sigma_z} + O(\delta^\frac{3}{2}).
\end{equation}
This path is not itself a geodesic, but it doesn't need to be to upper-bound the complexity distance
\begin{equation}
\mathcal{C}[e^{i \sigma_z \delta } ] \leq 2 \sqrt{\delta}. \label{eq:easyroutelocus}
\end{equation}
The reason for the square-root growth is that this path only generates $\sigma_z$ to the extent that $\sigma_x$ and $\sigma_y$ fail to commute; in the language of differential geometry, the failure of parallel transport is proportional not to the perimeter of the path but to the area. (We saw the same square-root growth, and for the same reason, in the state geometry of Eq.~\ref{eq:distancesimplestcaseequator}.)

For very small $\delta$, the direct  geodesic is always the shortest. However beyond a certain point, known as the `cut locus' \cite{cutlocusonbergersphere}, the indirect path becomes cheaper. 
For large $\mathcal{I}_{zz}$, the (first-order) transition between Eq.~\ref{eq:hardroutelocus} and Eq.~\ref{eq:easyroutelocus} occurs at
\begin{equation}
\delta_\textrm{cut locus} \sim \frac{1}{{\mathcal{I}_{zz}}}  \ \ \ \longleftrightarrow \ \ \  \mathcal{C}_\textrm{cut locus} \sim \frac{1}{\sqrt{{\mathcal{I}_{zz}}}}. \label{eq:cutlocusdirection}
\end{equation}
After going much less than a single unit of complexity in the $\sigma_z$-direction, it is already cheaper to take a `circuitous' route round the easy directions. To use a loose nautical metaphor, rather than sail directly into the wind we tack back and forth.\footnote{We can repeat this for the completely anisotropic case $\mathcal{I}_{zz} \gg \mathcal{I}_{yy} \gg \mathcal{I}_{xx} =1$. 
As before, the direct path just ploughs along in the $\sigma_z$-direction and has complexity given by Eq.~\ref{eq:hardroutelocus} as $\mathcal{C}[\textrm{direct}] = \sqrt{\mathcal{I}_{zz}} \delta$. The indirect path uses that 
\begin{equation}
e^{i \sigma_z \delta} \sim e^{i \sigma_x \theta_x} e^{i \sigma_y \theta_y} e^{-i \sigma_x \theta_x} e^{-i \sigma_y \theta_y}
\end{equation}
so long as the area of the rectangle is large enough $\theta_x \theta_y \sim \delta$ and $\theta_x, \theta_y \ll 1$. The length of this indirect path is $\mathcal{C}[\theta_x, \theta_y] \sim \theta_x + \sqrt{\mathcal{I}_{yy}} \theta_y$, and minimizing this subject to the constraint $\theta_x \theta_y \sim \delta$ gives 
\begin{equation}
\mathcal{C}[\textrm{indirect}]  =  \mathcal{I}_{yy}^{{1}{/4}} \sqrt{\delta}.
\end{equation}
The cut locus occurs when the direct and indirect paths have the same length
\begin{equation}
\delta_\textrm{cut locus}  \sim  \frac{\sqrt{\mathcal{I}_{yy} }}{\mathcal{I}_{zz}} \ \ \leftrightarrow \ \  \mathcal{C}_\textrm{cut locus} \sim \sqrt{\frac{\mathcal{I}_{yy} }{\mathcal{I}_{zz}}}.
\end{equation}} 

Indeed, we already saw this prefigured in Sec.~\ref{sec:whynegativecurvature} where negative curvature means that rather than proceeding directly down the hypotenuse at 45$^\circ$ it is cheaper to hew close to an easy direction down one arm of the right-angled triangle, and then switch to the other easy direction for the second arm. It is a general feature of complexity geometry---both for a single qubit and also for the many qubit case---that the cheapest way to move a long distance in a hard direction is to tack back and forth in the easy directions. \\

In the limit $\mathcal{I}_{zz} \rightarrow \infty$, the cut locus is driven towards the origin, and so the linear regime of Eq.~\ref{eq:hardroutelocus} shrinks to zero.  The unitary complexity in the difficult direction is given by \cite{ComplexityOfeisigmaz}
\begin{equation}
\mathcal{C}[e^{i \sigma_z \delta } ] \biggl|_{\mathcal{I}_{zz} = \infty, \ \mathcal{I}_{xx} = \mathcal{I}_{yy} =1} = \sqrt{ \delta (2 \pi - \delta) }.  \label{eq:thisreferanthere}
\end{equation}
This is analogous to the state-complexity result in Eq.~\ref{eq:distancesimplestcaseequator}. \\

Starting at any point, and proceeding along any geodesic, one eventually arrives at a cut locus. The distance to the cut locus depends on the hardness of the geodesic. Equation~\ref{eq:cutlocusdirection} shows that in \emph{hard} directions, the cut locus is close. By contrast, in \emph{easy} directions, the cut locus is much farther away:  
  to make $e^{i \sigma_x \delta}$, the cheapest way is generally just to proceed directly in the (easy) $\sigma_x$-direction. It is only for $\delta \geq \pi$---which is to say only once that antipode at $- \mathds{1}$ has been reached---that the cut locus is encountered. 

\subsection{Close in inner product, far in complexity} \label{sec:closeinnerfarcomplexity}

The example of Fig.~\ref{fig-NearEarth} shows two states that are maximally separated in inner-product distance but relatively close in complexity distance. The opposite situation can also arise: we can have two states that are close in inner product but far in complexity. Let's examine three aspects of the relationship between inner-product distance and complexity distance; we will see that the single-qubit example captures only the first two. 

\begin{enumerate}
\item Two states that are arbitrarily close in inner product will be arbitrarily close in complexity, and vice versa.  
\end{enumerate}
The complexity metric and inner-product metric are topologically identical, which means if two points are arbitrarily close in one metric they are arbitrarily close in the other. More specifically,  two states that are separated by an inner-product distance $\delta$ have a complexity distance bounded by 
\begin{equation}
\sqrt{\mathcal{I}_\textrm{min}} \delta \leq \Delta \mathcal{C} \leq \sqrt{\mathcal{I}_\textrm{max}} \delta, \label{eq:trappedcomplexity}
\end{equation}
where  $\mathcal{I}_\textrm{min}$ and $\mathcal{I}_\textrm{max}$ are the smallest and largest penalty factors. 
That the geometric definition of complexity is a continuous function of SU($2^N$) and $\mathbb{CP}^{2^N-1}$ is in marked contrast to the gate definition of complexity outlined in Sec.~\ref{sec:gatecomplexity}. An immediate corollary of Eq.~\ref{eq:trappedcomplexity} is that if $\mathcal{I}_\textrm{min} \geq 1$ then the complexity distance is never less than the inner-product distance. 

\begin{enumerate}
\item[2.] The complexity distance can be a huge multiple of the inner-product distance. 
\end{enumerate}
In the metric of Eq.~\ref{eq:BergerMetric} (and Eq.~\ref{eq:statespacecomplexitymetric}) the ratio of the complexity distance to the inner-product distance can be very large. Indeed, for small enough $\delta$ the upperbound of Eq.~\ref{eq:trappedcomplexity} is tight, 
\begin{equation}
\textrm{for } \delta \, \, \, \lsim \, \, \, {\mathcal{I}_{zz}^{ \, - 1} } : \ \ \ \ \ \ \ \ \frac{ \mathcal{C} [ e^{i \sigma_z \delta} ]}{\delta}   = \sqrt{\mathcal{I}_{zz}}.
\end{equation}
Even though two points are separated by only a small inner-product distance, if the direct route passes in a hard direction then the complexity distance gets multiplied by a huge number, and this ratio becomes ginormous. 

\begin{enumerate}
\item[3.] The complexity distance can be huge even while the inner-product distance is small.

(True for the large-$N$ complexity metric; \emph{not} true for the single-qubit complexity metric.)
\end{enumerate}
In the complexity metric with a large number $N$ of qubits discussed in Sec.~\ref{sec:complexitygeometry}, there may be states that are exponentially close (in $N$) in inner product but exponentially far (in $N$) in complexity. However,  this is a feature that is \emph{not} captured by the single-qubit example with $\mathcal{I}_{zz} \gg \mathcal{I}_{xx} = \mathcal{I}_{yy} = 1$. As we will now show, in the single-qubit metric all complexity distances are upperbounded by an $O(1)$ number no matter how large $\mathcal{I}_{zz}$ gets:

\begin{itemize}
\item[$\circ$]{One-qubit unitary complexities are never large.} 

For the \emph{inner-product}  metric, the two unitaries $\mathds{1}$ and $- \mathds{1} = \exp [ i {\pi} \sigma_z]$ are as distant as two unitaries can be: they lie at opposite antipodes of the $S^3$ and are therefore separated by  $\pi$. 
You might imagine that turning on a large penalty factor $\mathcal{I}_{zz}$ would greatly increase the distance between these two unitaries, since you are punishing movement in the $\sigma_z$-direction. But in fact increasing $\mathcal{I}_{zz}$ doesn't increase the distance \emph{at all}, since there remains a path in the unpunished $\sigma_x$-direction using $- \mathds{1} = \exp [ i {\pi} \sigma_x]$. 

Generalizing away from the antipodal case, it is clear that there are no pairs of points that become very distant since any SU(2) can be written in `proper' Euler coordinates as $e^{i \sigma_x \theta_1} e^{i \sigma_y \theta_2} e^{i \sigma_x \theta_3}$. Indeed, Ref.~\cite{diameterofbergersphere} shows that no matter how large $\mathcal{I}_{zz}$, no distance exceeds $\pi$, 
\begin{equation}
0 \leq \textrm{inner-product distance} \leq \textrm{complexity distance}\Bigl|_{\mathcal{I}_{zz} \geq \mathcal{I}_{xx} = \mathcal{I}_{yy} = 1}   \leq \, \pi.
\end{equation}

\item[$\circ$]{One-qubit state complexities are never large.} 

The same pattern applies to state complexity, considered in Sec.~\ref{sec:stateC}. For the inner-product metric on quantum states the maximum distance is $\pi/2$. Inspecting Eq.~\ref{eq:statespaceIinfinity} shows that making $\mathcal{I}_{zz}$  arbitrarily large  does not increase the maximum distance
\begin{equation}
0 \leq \textrm{inner-product distance} \leq \textrm{complexity distance}\Bigl|_{\mathcal{I}_{zz} \geq \mathcal{I}_{xx} = \mathcal{I}_{yy} = 1}  \leq \, \frac{\pi}{2}.
\end{equation}
\end{itemize}

In summary, in the single-qubit example adding a large penalty factor may greatly multiply the distance between two points that start off \emph{close}, but does not greatly increase the distance between two points that start off already \emph{distant}. The explanation is related to a phenomenon discussed in Sec.~\ref{sec:cutloci}: sufficiently distant points will generally be beyond the cut locus, and therefore connected together by many different geodesics. Some geodesics will be short (in inner-product distance) but hard (large $\mathcal{I}$), others will be long (in inner-product) but easy (small $\mathcal{I}$). The easier paths will be mostly unaffected by making $\mathcal{I}_{zz}$ arbitrarily large. The reason that there is still a relatively short easy path is that for the single-qubit example a large fraction of the  directions are easy---two out of three. This is very different to the multiqubit case of Sec.~\ref{sec:complexitygeometry} where easy directions are by assumption exponentially rare. 

(Of course, we \emph{can} make the single-qubit example capture that complexities can be large by making  \emph{two} of the moments of inertia large\footnote{Reference \cite{diameterofbergersphere} shows that for $\mathcal{I}_{xx} = \mathcal{I}_{yy} > 2 \mathcal{I}_{zz}$, the maximum distance is $\pi \mathcal{I}_{xx} /  \sqrt{\mathcal{I}_{xx} - \mathcal{I}_{zz}}$.}, or even more trivially and even more tastelessly by making all \emph{three} large. But even if we did, the single-qubit complexity geometry would still fail to capture \emph{why} the complexity can grow so large, and in particular would fail to exhibit the fractal nature of the complexity frontier described in Sec.~\ref{sec:notgeneral}.)

     \begin{figure}[htbp] 
    \centering
    \includegraphics[width=3in]{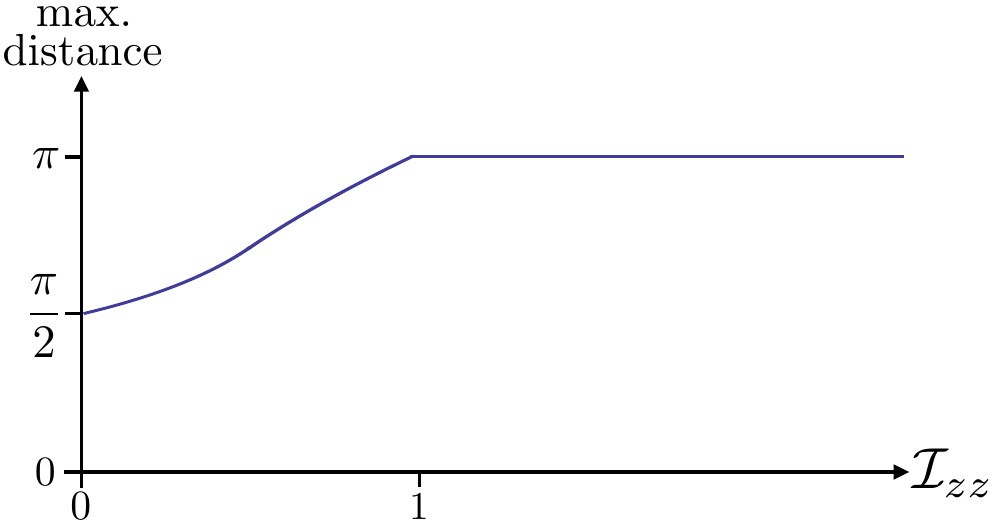} 
    \caption{The maximum separation on the Berger sphere as a function of $\mathcal{I}_{zz}$, for $\mathcal{I}_{xx} = \mathcal{I}_{yy} = 1$. The maximum distance is never greater than $\pi$, no matter how large $\mathcal{I}_{zz}$; on the other hand the maximum distance never goes below $\pi/2$ no matter how small $\mathcal{I}_{zz}$  \cite{diameterofbergersphere}.}
 \end{figure} 
 \FloatBarrier

 Even though the maximum distance does not grow as $\mathcal{I}_{zz}$ gets large, there are two simple geometrical quantities that do. The first is the volume---the volume of the Berger sphere depends linearly on $\mathcal{I}_{zz}$---and indeed for the general higher-dimensional complexity metric the volume also gets large as the penalty factors grow. The second quantity is the \emph{average} distance. As $\mathcal{I}_{zz}$ grows, the average distance between two points on the Berger sphere gets closer to the maximum distance. It is also a property of the higher-dimensional complexity geometry that the average distance is close to the maximum distance.

\subsection{Complexity geometry \emph{vs.} gate complexity} \label{section4:complexityvsgate}

The single-qubit example accurately illustrates some of the relative merits of the geometric and gate-counting definitions of complexity. \\

In this paper we have discussed two different definitions of computational complexity: gate complexity (Sec.~\ref{sec:gatecomplexity}) and complexity geometry (Sec.~\ref{sec:complexitygeometry}). That there are multiple definitions is to be expected: computational complexity measures how hard a given transformation is to implement, and so it is no surprise that two different people with two different sets of abilities will disagree about how difficult is a given task. We are now in a position to compare these two definitions. 

 \begin{center}
\begin{tabular}{c||c} 
GATE COMPLEXITY \  \  & \ \   COMPLEXITY GEOMETRY  \Bo \\
\hline
\hline
complexity is number of & complexity is length of   \To\\
  gates in smallest circuit that & the shortest path that   \\
  implements transformation & implements transformation
   \Bo \\
\hline
only allowed steps are  &  all directions allowed but  \To  \\ 
primitive gates &  some   highly penalized 
\Bo  \\
\hline
primitive gates typically  &  $k$-local directions typically  \To  \\ 
taken to be $k$-local  & assigned low penalty (unpenalized) \Bo  \\
\hline
tolerance $\epsilon$ required when  &  no tolerance $\epsilon$ required: \To  \\ 
 primitive gate set is countable &  can hit target exactly \Bo  \\
\hline
complexity is discontinuous & complexity is continuous    \To  \\ 
function of SU($2^N$) \& $\mathbb{CP}^{2^N-1}$ & function of SU($2^N$) \& $\mathbb{CP}^{2^N-1}$  \Bo  
\end{tabular}
\end{center}

In Sec.~\ref{sec:gatecomplexity} we saw some of the shortcomings of the gate definition of complexity when applied to systems undergoing evolution by a continuous Hamiltonian. Let's see how the complexity geometry definition fixes or ameliorates those shortcomings. 

First, in the gate definition the choice of which primitive gates are permitted, while perhaps dictated by real-world engineering constraints, seems arbitrary. Why those particular $k$-local gates and not others? In the complexity-geometry definition, all possible $k$-local directions (i.e. all terms in the Hamiltonian that are the tensor product of at most $k$ one-qubit $\sigma_i$s) are permitted, and typically with the same penalty factor too. The only arbitrariness that remains is deciding how severely to punish non-$k$-local directions as a function of their non-$k$-locality (deciding what schedule of penalty factors $\mathcal{I}_{ij}$ to apply as a function of the weight).

Second, the gate definition of complexity as described in Sec.~\ref{sec:gatecomplexity} required the introduction of a `tolerance' $\epsilon$, which described how close we must come to the target state before declaring victory. This is no longer necessary with complexity geometry: movement in all directions in Hilbert space is now permitted (even if punished by a large penalty factor), and every target state can now be hit exactly. 

Third, the gate definition of complexity is discontinuous in the sense that two points may be arbitrarily close in inner product but still have very different complexities. By contrast, Eq.~\ref{eq:trappedcomplexity} told us that,  when the penalty factors are all finite, points that are arbitrarily close in inner product are also arbitrarily close in the complexity metric, so the map from inner-product geometry to complexity geometry is continuous. \\

What exactly is it about complexity geometry that allows it to dispense with a tolerance and that makes it continuous? Examining Eq.~\ref{eq:trappedcomplexity}, you might imagine that the essential feature is that the complexity geometry permits direct motion in \emph{every} direction. But that's overkill. Even if direct motion in some directions were completely forbidden---even if the penalty factors in some directions were infinite---then so long as a generating subset is still permitted we can build up everything by tacking back and forth and there is no need for a tolerance. Furthermore, since infinitesimal motion in the permitted directions has infinitesimal cost, the map from inner-product geometry to complexity geometry is continuous. Thus even if we took $\mathcal{I}_{zz} = \infty$ in the single-qubit complexity geometry of Eq.~\ref{eq:BergerMetric} (and even if we took $\mathcal{I}_{k=3} = \mathcal{I}_{k=4} = \ldots = \mathcal{I}_{k=N} = \infty$ in the multi-qubit complexity geometry), there would still be no need for a tolerance and the complexity geometry would still be continuous.\\

It is worth clarifying what `continuity'  does and does not mean. On the one hand we have just seen that the complexity metric is continuous. On the other hand we discussed in Sec.~\ref{sec:closeinnerfarcomplexity} that  in the multi-qubit example there are pairs of points that are exponentially close in inner product but exponentially far in complexity. There is no contradiction because continuity is a statement about \emph{infinitesimals}. On short enough scales the complexity doesn't vary much from point to point---it is clear that for $\delta < \mathcal{I}_{\textrm{max}}^{\ \ -\frac{1}{2}}$ the complexity varies by less than a single unit, and in the last paragraph we argued that the existence of indirect easy paths means the variation in complexity may be small even for  somewhat larger $\delta$. But move too far and the complexity starts varying dramatically. If for some experimental reason we were constrained to measure the complexity with an inner-product resolution worse than the characteristic scale of variation, the complexity distance would be functionally discontinuous. \\

We have highlighted three big differences between complexity geometry and the gate definition of complexity. Let's mention one more. When compiling circuits out of a countable set of primitive gates, at each step motion is restricted to discrete jumps in a discrete set of directions. This means  there are many fewer options than when applying an arbitrary Hamiltonian, and this limitation gives rise to a pattern of gate complexity that is considerably more intricate than that produced by the complexity geometry. As one example, consider the set of states that have a complexity below some value: in the complexity geometry this set forms a single connected component; in the gate definition the set breaks up into numerous disconnected islands. As another example, we have seen that the complexity geometry of a single qubit is too simple to generate large complexity; by contrast the gate complexity of a single qubit may be huge. 

(Indeed, when only a discrete set of gates are permitted, even the complexity of just a single U(1) phase is rich enough to be interesting. Fernando Pastawski has explained to us that if we are are permitted to pick only one gate, the choice that minimizes the average complexity of U(1) is the `golden gate' that rotates by the Golden Ratio. This is because the Golden Ratio optimally avoids near-collisions by having the `most irrational' continued fraction expansion.)\\

For some applications, the so-called `shortcomings' of the gate definition of complexity may actually be advantageous. For other applications, we might try to modify the gate definition so that it becomes more like the complexity geometry and inherits some of its features \cite{Jefferson:2017sdb,Chapman:2017rqy,Cottrell:2017ayj,Nielsen1,Nielsen2,Nielsen3,Nielsen4,Khaneja}. But for our intended applications, the complexity geometry definition seems superior.

\subsection{What $N=1$ fails to capture} \label{sec:notgeneral}
We have seen that the single-qubit example manifests many of the important features of complexity geometry. What are the features that the single-qubit case does \emph{not} capture?  

The short answer is that the single-qubit example fails to capture anything that relies on the distinction between scaling polynomially, exponentially, or double-exponentially with $N$. The difference between these scalings gets washed out when $N=1$. This means that the single-qubit example misses many of the `statistical' properties of complexity. 

For example, for the general-$N$ complexity geometry described in Ref.~\cite{Nielsen1} and in Sec.~\ref{sec:complexitygeometry}, there are vastly more hard directions than easy directions: exponentially many hard directions versus only polynomially many $k$-local directions. By contrast, for the single-qubit example we focussed on in this paper, there are more easy directions than hard: two versus one. 

Similarly, for the general-$N$ case there are double-exponentially more high-complexity states than low-complexity states, with the consequence that there may emerge a statistical thermodynamics of complexity, and in particular an analog of the Second Law \cite{Brown:2017jil}. By contrast, for the single-qubit case, there is no parametric separation between the time to reach maximum complexity [O$(2^N)$] and the quantum recurrence time [O$(2^{2^N})$]. 

Another feature of large-$N$ complexity geometry that the single-qubit complexity geometry fails to capture is the fractal-like structure of the set of low complexity states. Consider the set of $N$-qubit unitaries with a complexity  less than some value. For very small values, this set is a high-dimensional ball around the identity, bounded by a spherical `complexity frontier'. Measured in the complexity metric this ball looks round; measured in the inner-product metric the ball is deformed by being stretched in easy directions and squeezed in hard directions. For larger values of the complexity, the set becomes increasingly convoluted, with multidimensional tendrils wrapping many times round the SU($2^N$) in easy directions while making little progress in the hard directions. The tendrils are  intertwined and give rise to a fractal-like structure. The topology of the set becomes increasingly complicated as it intersects itself at cut loci. We expect the one-qubit complexity geometry to be too simple to exhibit this convoluted behavior\footnote{Though it might be interesting to numerically investigate the complexity geometry for the case $\mathcal{I}_{zz} \gg \mathcal{I}_{yy} \gg \mathcal{I}_{xx}$ to see what vestiges remain.}.

\section{Conclusion} \label{sec:conclusion}
\sc
 In complexity geometry, nothing is ever forbidden, it is merely expensive. Rather than leaping through Hilbert space in discrete gate-sized jumps, now steps of any size and in any direction are permitted. However, steps in difficult directions are penalized in proportion to their difficulty. In this paper we discussed why this geometric definition of complexity is one that is well-suited to  holographic applications, and have studied the single-qubit example as a simple tractable case that nevertheless embodies many of the general phenomena that apply to complexity geometries in general. \\

 There is prior literature on using single-qubit examples as simple laboratories to study gate complexity. For a very carefully chosen set of primitive gates, it is possible to used advanced mathematics to calculate the gate complexity efficiently \cite{SingleQubitGates}. For more general sets of primitive gates, Solovay and  Kitaev proved a theorem about how accurately one-qubit operators can be approximated \cite{SolovayKitaev}. As discussed in Sec.~\ref{section4:complexityvsgate}, since the gate definition of complexity limits evolution to discrete jumps, the resultant pattern of complexity is more intricate and convoluted.

There is also prior literature on seeking simple models of complexity geometry. 
The complexity geometry has been explored for harmonic oscillators \cite{Jefferson:2017sdb}, coherent states \cite{Guo:2018kzl}, free fermions \cite{Hackl:2018ptj}, the permutation group \cite{Lin:2018cbk}, and other systems \cite{Chapman:2017rqy,Yang:2017nfn,Yang:2018cgx,NotInInspires2013,Bhattacharyya:2018bbv,Chapman:2018hou,Khan:2018rzm,Caputa:2018kdj}.

There is even prior literature on considering the complexity geometry of a single qubit. In Ref.~\cite{NielsenSingleQubit}, Gu, Doherty, \& Nielsen considered the complexity geometry of SU(2) with, in our notation, $\mathcal{I}_{xx} = \mathcal{I}_{yy} = 1$ and $\mathcal{I}_{zz} = 0$. 
They showed that in this limit the metric of Eq.~\ref{eq:BergerMetric} becomes that of a round \emph{two}-sphere. Notice that this limit is  opposite to the one we focus on in this paper: we have argued that studying the limit $\mathcal{I}_{zz} \gg  \mathcal{I}_{xx} = \mathcal{I}_{yy}$ produces generalizable morals by better embodying the characteristic phenomena of  multi-qubit complexity geometry.\\

In this paper we have looked at complexity geometry with the smallest positive integer number of qubits: one. The next smallest possible integer number of qubits is two. Following the method of Sec.~\ref{sec:rigidbodies}, two-qubit unitary complexity geometry can be related to the rotation of rigid bodies in \emph{six} dimensions, since SU(4) $\sim$ SO(6). 
In the two-qubit case it would be tempting to make the two-local operators `hard' (large moments of inertia), and make the one-local operators `easy' (small moments of inertia). However, it would actually better reflect the large-$N$ case if the assignment were the other way round. In Sec.~\ref{sec:whynegativecurvature} we saw that to generate negative curvature, the commutator of two easy directions should be hard. But one-local operators commute to other one-local operators, so making the one-local directions easy will not generate negative curvature.  Instead, to faithfully imitate the large-$N$ complexity geometry, even if not to accurately reflect a realistic laboratory capability,  the one-local directions should be hard and it should be the two-local directions that are easy.

Another simple example would be the complexity geometry of the Heisenberg group. The Heisenberg group is in some sense even simpler than the single qubit, since one of the generators---the identity, which generates a global phase---commutes with the other two: $[\hat{\mathds{1}},\hat{x}] = [\hat{\mathds{1}},\hat{p}]=0$. The only non-trivial commutator is $[\hat{x}, \hat{p}] = i \hat{\mathds{1}}$. In order that easy directions commute to a hard direction, both $\hat{x}$ and $\hat{p}$ should be easy, and $\hat{\mathds{1}}$ should be hard.  \\

Even in the single-qubit case, there is a limit to what we have been able to wring from analytic analysis. Clear targets for numerical investigation include:   tracing the cut locus; calculating the average distance between two points; mapping the shape of what we have called the `complexity frontier'; and calculating the distance between two unitaries generated by evolving for a time $t$ with slightly different easy Hamiltonians (the `switchback' configuration). Given the relatively low dimensionality, this should be within computational reach for general $\mathcal{I}_{xx}$, $\mathcal{I}_{yy}$, and $\mathcal{I}_{zz}$, and indeed even if generalized to a handful of qubits. \\

In this paper we have examined the complexity of a single qubit. In the limit we focussed on---large $\mathcal{I}_{zz}$ but still moderate $\mathcal{I}_{xx}$ and $\mathcal{I}_{yy}$---the complexity never gets very large. When normalized so that a unit step in an easy direction reaches an orthogonal state, the largest relative complexity is never more than O(1). From the perspective of the gate definition of complexity it is somewhat surprising that increments of complexity less than one---``sub-gate complexity''---have any meaning. And yet they do: within a single unit of complexity we have found most of the key phenomena of computational complexity.  We even found negative curvature that leads to the exponential growth of perturbations, despite the fact that this phenomenon is commonly explained in terms of an epidemic spreading amongst the qubits \cite{Susskind:2014jwa}, an explanation that makes no sense whatsoever when there is just a single qubit.  When there are $N$ qubits, the corresponding surprise is that it is sensible to talk about increments of complexity less than $N$---a circuit depth less than one---and yet we have seen that from the perspective of complexity geometry there is nothing wrong with considering ``sub-depth-one complexity".

In the AdS/CFT correspondence, both the AdS (`bulk') side and the CFT (`boundary') side of the duality are separately `local', in the sense that the interactions in each theory arrange themselves so that only nearby points interact. From tensor network constructions of holographic theories it is clear how the locality of the CFT is directly inherited by the AdS so as to give rise to bulk-locality on scales longer than the AdS-length. But the AdS that arises from holographic field theories is local not merely on scales \emph{longer} than the AdS-length, but also on scales much \emph{shorter} than the AdS-length, all the way down to the string/Planck length. From the CFT point of view this is something of a surprise. From the CFT point of view it is not at all obvious that it makes sense to talk about bulk regions smaller than an AdS length, and it is not well understood how the CFT's strongly coupled matrix degrees of freedom arrange themselves so as to give rise to ``sub-AdS locality".

The holographic complexity conjecture relates the computational complexity of the CFT to the volume or action of regions of the AdS bulk \cite{Susskind:2014rva,Stanford:2014jda,Brown:2015bva,Brown:2015lvg}. In light of the forgoing it is interesting that the `surprise' that locality makes sense even on sub-AdS scales  is mapped by the duality onto the `surprise' that complexity makes sense even on sub-depth-one scales. A single AdS volume in the complexity-volume conjecture \cite{Stanford:2014jda}, and single AdS-sized Wheeler-de-Witt patch in the complexity-action version of the conjecture \cite{Brown:2015bva,Brown:2015lvg}, gets mapped to a complexity equal to the dimension of the boundary gauge group, which is to say a complexity equal to the number of qubits---a depth-one circuit. If the holographic complexity conjecture is to resolve sub-AdS scale features of the bulk, it is necessary to have a definition of complexity that can resolve circuit-depths of less than one. The ultimate limit of locality---a single Planck-sized region---corresponds to a complexity far less than a single gate.\\


In this paper we have explored different definitions of `complexity'. Ultimately, the question of which definition is the best is really a question of which definition is the most useful, which in turn depends on what system we wish to study, and to what purpose.  For us, the most interesting definition of complexity would be the one holographically dual to the size of the black-hole wormhole. With this motivation, we investigated the complexity geometry definition. We did this by calculating the complexity geometry of a single qubit. Within a single qubit we found many of the signature phenomena of the full multi-qubit complexity geometry: we saw that the geometry is right-but-not-left invariant, and saw the impact that has on  homogeneity and isotropy; we saw that geodesics  are no longer generated by time-independent Hamiltonians; we saw that easy-easy sections develop negative curvature, and saw the consequences that has for the exponential growth of perturbations; we saw that highly anisotropic penalty factors drive the cut locus close to the origin, and saw the repercussions this has on the pattern of complexity distances; we saw that the ratio between the complexity distance and the inner-product distance may be ginormous; and we saw exhibited the relative merits of the gate-counting and geometric definitions of complexity. In short, we saw that the humble Bloch sphere, when squashed by anisotropic penalty factors, prefigured and illustrated many of the features of the complexity geometry of an exponentially larger Hilbert space.

\section*{Acknowledgements}
We thank Michael Freedman, Henry Lin, and Ying Zhao for helpful discussions.  This publication was made possible in part through the support of a grant from the John Templeton Foundation and by NSF Award Number 1316699.
\appendix

\section{Journey to the Center of the Earth} 

 To demonstrate the unity of physics, let's extraneously point out an entirely different application of the $\mathcal{I}_{zz} = \infty$ single-qubit state complexity geometry.
\subsection{Complexity geometry of a single infinitely-squashed qubit} 
In Sec.~\ref{subsec:infiniteIzz} we looked at the infinitely squashed  Bloch sphere. This is the complexity geometry of single-qubit {states} when $\mathcal{I}_{xx} = \mathcal{I}_{yy} = 1, \, \mathcal{I}_{zz} = \infty$. Even though you are not allowed to move directly in the $\sigma_z$ direction, you can still synthesize $\sigma_z$ indirectly by zig-zagging in the $\sigma_x$ and $\sigma_y$ directions. If we parametrize single-qubit states by 
\begin{equation}
| \psi \rangle = \cos \frac{\theta}{2} \,  |0 \rangle + \sin \frac{\theta}{2} \,  e^{i \phi} |1 \rangle \ , 
\end{equation}
then Eq.~\ref{eq:statespaceIinfinity} gave the metric on the infinitely-squashed Bloch sphere as 
\begin{eqnarray}
4 ds^2\biggl|_{\mathcal{I}_{zz} = \infty, \ \mathcal{I}_{xx} = \mathcal{I}_{yy} =1} &=& d \theta^2 + \tan^2 \theta d \phi^2. \label{eq:metrictobereferredto123}
\end{eqnarray}
In this metric, the distance between two points both on the equator is given by Eq.~\ref{eq:distancesimplestcaseequator} as
\begin{equation}
\textrm{geodesic:} \ \ \ \  \ \Delta s \Bigl|_{\theta_1 = \theta_2 = \frac{\pi}{2} , \, \mathcal{I}_{zz} = \infty, \, \mathcal{I}_{xx} = \mathcal{I}_{yy} = 1} = \frac{1}{2} \sqrt{ \Delta \phi (2 \pi - \Delta \phi)} \  . \label{eq:A3}
\end{equation}
As discussed in Sec.~\ref{sec:geoHilbertvsH}, 
the geodesic will be described by a time-\emph{dependent} Hamiltonian of the form $H(t) = a(t) \sigma_x + b(t) \sigma_y$. If instead we insist on connecting two points on the equator by a time-\emph{independent} Hamiltonian, then the optimal fixed rotation axis pierces the equator halfway between the two points, and the distance is 
\begin{equation}
\textrm{fixed $H$:} \ \ \ \ \ \ \ \ \ \ \ \ \ \ \ \Delta s \Bigl|_{\theta_1 = \theta_2 = \frac{\pi}{2} , \, \mathcal{I}_{zz} = \infty, \, \mathcal{I}_{xx} = \mathcal{I}_{yy} = 1} = \frac{\pi}{2},  \label{eq:fixedHamiltonianfixedtime}
\end{equation}
 for any $\Delta \phi$. Time-independent Hamiltonians do not, in general, give geodesics. \\

\noindent Notice that for $\Delta \phi = \pi$, the following three coincide: the geodesic distance with $\mathcal{I}_{zz} = \infty$, the geodesic distance with $\mathcal{I}_{zz} = 1$, and the time-independent Hamiltonian distance.  
\subsection{Ground shipping} 
A common sci-fi fantasy (beloved of those who know more about physics than geology) is that we might one day deliver intercontinental mail using tunnels bored clear through the Earth. The beauty of this scheme is that no propulsion is required, since the Earth's gravitational field both accelerates the package at the start of its journey, and decelerates the package at the end. 

At the risk of getting diverted by boring logistics, we can ask for the optimal shape for this tunnel. The shortest \emph{in space} tunnel is a straight line. But we don't care how far the mail travelled, we care how soon it arrives. The deeper the package plunges, the faster it goes, and so the \emph{shortest-in-travel-time} tunnel  delves deeper than a straight line. Were the Earth flat, the minimal-time tunnel---the brachistochrone---would be a cycloid. Since the Earth is round, to derive the optimal shape will need a calculation that surpasses Newton's. 

Assuming the Earth is a sphere of radius $R$ and constant density $\rho$ and ignoring spin and friction, conservation of energy tells us that 
 the velocity of a parcel that starts at rest at the surface is 
\begin{equation}
 v(r)^2 = \frac{4 \pi G \rho}{3} (R^2 - r^2) \ . 
\end{equation}
Let $\Delta \phi$ be the angle between the two points, as measured from the center of the Earth. Defining $r  = R \sin \psi$, the incremental travel time is 
\begin{equation}
dt^2 = \frac{dr^2 + r^2 d \phi^2 }{v(r)^2} = \frac{dr^2 + r^2 d \phi^2 }{4 \pi G \rho (R^2 - r^2)/3 } = \frac{d \psi^2 + \tan^2 \psi d \phi^2}{4\pi G \rho/3}  \ .
\end{equation}
Up to a rescaling, this is the same as Eq.~\ref{eq:metrictobereferredto123}.

     \begin{figure}[htbp] 
    \centering
    \includegraphics[width=4.5in]{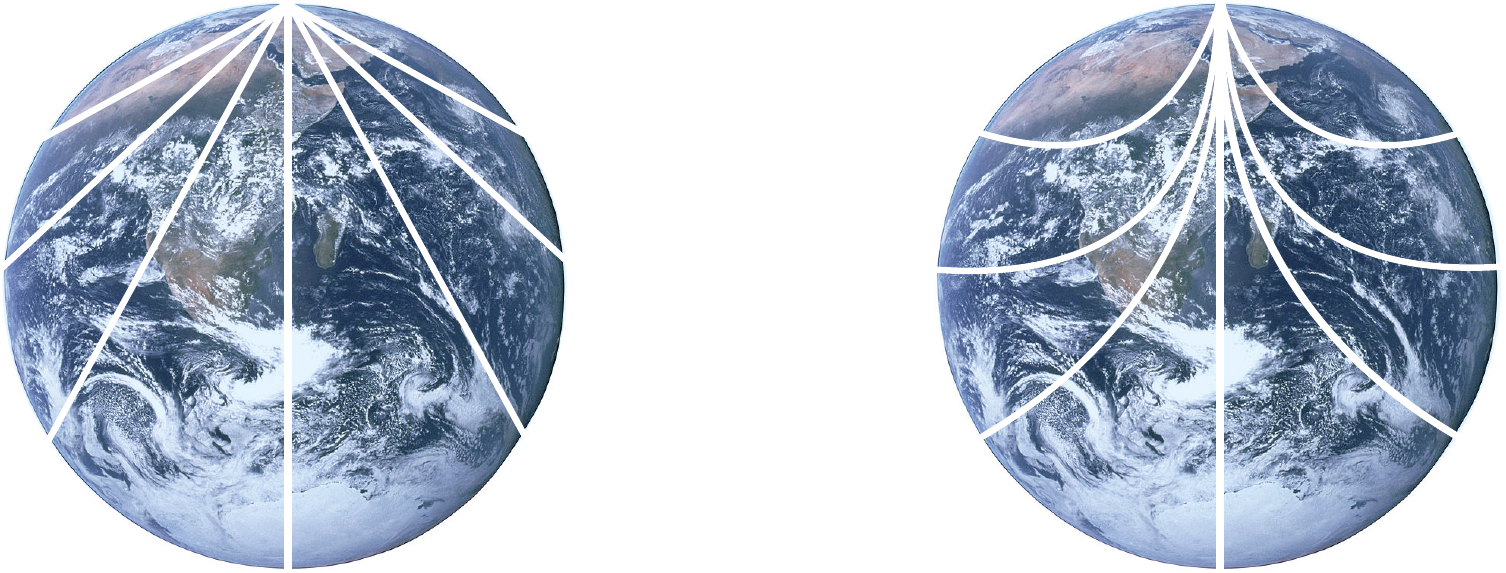} 
    \caption{Globe and mail: two strategies for boring mail-delivery tunnels through the Earth. Left: we bore straight-line tunnels. In this case, mail inserted at one end of a tunnel arrives at the other end a time $\Delta t =  \sqrt{{3 \pi }/{4  G \rho}}$ later, no matter which tunnel it takes. (This is {also} the time it takes to complete a half-orbit at zero altitude.)  Right: expedited delivery is achieved by boring curved tunnels, which follow the geodesics of the sub-Riemmanian single qubit complexity geometry Eq.~\ref{eq:metrictobereferredto123}.}
    \label{fig:journeytothecenter}
 \end{figure}

\subsection{Quantum tunneling}
Let us make explicit the connection between quantum complexity geometry and tunneling through the Earth.  
\begin{itemize}
\item Fixed Hamiltonian $\leftrightarrow$ straight tunnel. 

In Eq.~\ref{eq:fixedHamiltonianfixedtime} we saw that to connect two states on the equator of the Bloch sphere with a \emph{fixed} Hamiltonian took the same complexity distance no matter the $\Delta \phi$. 

The exact analogue of a fixed Hamiltonian is a straight-line tunnel. The transit time for a straight-line tunnel is 
\begin{equation}
\Delta t \Bigl|_\textrm{straight tunnel} =  \sqrt{\frac{3}{4 \pi G \rho}} \ \pi \ . 
\end{equation}
All straight-line tunnels have the same transit time:  when $\Delta \phi$ is small, the straight-line tunnel is short but slow. Indeed, not only is the straight-line transit time independent of $\Delta \phi$, it's also independent of $R$. 
\item Geodesic of complexity geometry $\leftrightarrow$ optimal tunnel. 

The exact analogue of a geodesic of the complexity geometry is a shortest-time tunnel. The transit time of a shortest-time tunnel is 
\begin{equation}
\Delta t \Bigl|_\textrm{fastest tunnel} =\sqrt{\frac{3}{4 \pi G \rho}} \ \sqrt{ \Delta \phi (2 \pi - \Delta \phi)}   \ . 
\end{equation}
This is the same (up to rescaling) as Eq.~\ref{eq:A3}. 

Since the analogy is exact, it's not just the transit times that agree, but also the shape of the curves. Project the geodesics of the Bloch sphere down onto the disc that cuts through the equator and they'll match exactly onto the tunnels in Fig.~\ref{fig:journeytothecenter}. 

The geodesics between points not on the equator of the Bloch sphere correspond to the optimal tunnels for deliveries that start with non-zero velocity. They are described by geodesics of the same metric, just with different initial conditions. 
\end{itemize}

\noindent This is an exact duality between a non-gravitational quantum system (the Bloch sphere), and a classical gravitational system (tunnels through the Earth).  

\end{document}